\documentclass[preprint,journal]{vgtc}                     

\onlineid{1368}

\vgtccategory{Research}

\vgtcpapertype{application/design study}

\title{SLInterpreter: An Exploratory and Iterative Human-AI 
Collaborative System for GNN-based Synthetic Lethal Prediction}

\author{%
  \authororcid{Haoran Jiang}{0009-0009-5717-4208},
  \authororcid{Shaohan Shi}{0009-0004-3384-8304},
  \authororcid{Shuhao Zhang}{0009-0008-1933-1869}, 
  \authororcid{Jie Zheng}{0000-0001-6774-9786} and 
  \authororcid{Quan Li}{0000-0003-2249-0728}
}

\authorfooter{
H. Jiang, S. Shi, S. Zhang, J. Zheng and Q. Li (corresponding author) are with School of Information Science and Technology, ShanghaiTech University, and Shanghai Engineering Research Center of Intelligent Vision and Imaging, China. E-mail: \{jianghr2023, shishh2023, zhangshh, zhengjie, liquan\}@shanghaitech.edu.cn.

}
\abstract{Synthetic Lethal (SL) relationships, though rare among the vast array of gene combinations, hold substantial promise for targeted cancer therapy. Despite advancements in AI model accuracy, there is still a significant need among domain experts for interpretive paths and mechanism explorations that align better with domain-specific knowledge, particularly due to the high costs of experimentation. To address this gap, we propose an iterative Human-AI collaborative framework with two key components: \textit{1) Human-Engaged Knowledge Graph Refinement based on Metapath Strategies}, which leverages insights from interpretive paths and domain expertise to refine the knowledge graph through metapath strategies with appropriate granularity. \textit{2) Cross-Granularity SL Interpretation Enhancement and Mechanism Analysis}, which aids experts in organizing and comparing predictions and interpretive paths across different granularities, uncovering new SL relationships, enhancing result interpretation, and elucidating potential mechanisms inferred by Graph Neural Network (GNN) models. These components cyclically optimize model predictions and mechanism explorations, enhancing expert involvement and intervention to build trust. Facilitated by \textit{SLInterpreter}, this framework ensures that newly generated interpretive paths increasingly align with domain knowledge and adhere more closely to real-world biological principles through iterative Human-AI collaboration. We evaluate the framework's efficacy through a case study and expert interviews.}


\teaser{
  \centering
  \includegraphics[width=\linewidth]{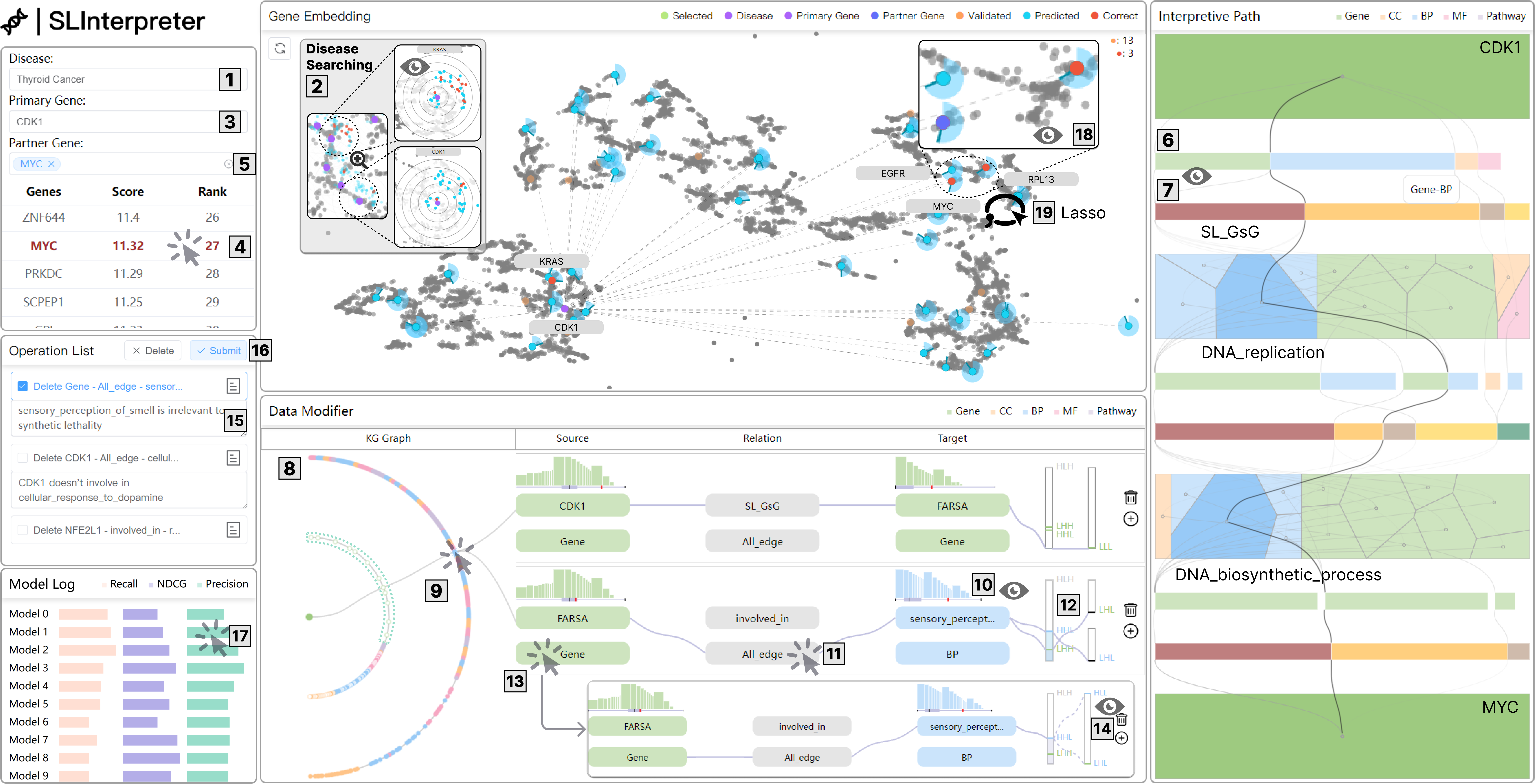}
  \vspace{-7mm}
  \caption{In Case Study, the biologist’s analysis comprises: \protect\caserectangle{1} Select the \textit{Thyroid Cancer} in the \textit{Disease search box}. \protect\caserectangle{2} Observe the similar pattern between KRAS and CDK1 by zooming in on the \textit{Radar} in the \textit{Embedding View}. \protect\caserectangle{3} Select CDK1 in the \textit{Primary Gene search box}. \protect\caserectangle{4} Select MYC, a correctly predicted partner gene of CDK1, in \textit{Partner Gene Table}. \protect\caserectangle{5} MYC tagged in \textit{Partner Gene search box}. \protect\caserectangle{6} The \textit{Entity Flow Bar} indicates that gene-gene connections are no longer predominant. \protect\caserectangle{7} The \textit{Path Bar} displays a smaller ratio of SL\_GsG. \protect\caserectangle{8} Narrow the investigation scope to \textit{CDK1$\rightarrow$Gene$\rightarrow$BP} by color encoding in the \textit{Knowledge Graph}. \protect\caserectangle{9} Select the irrational path in the \textit{Knowledge Graph}. \protect\caserectangle{10} Observe the histogram of \textit{sensory\_perception\_of\_smell}, indicating that this entity often ends such metapaths. \protect\caserectangle{11} Select \coloredboxLLH{L-H-L} strategy by clicking on \textit{All edge}. \protect\caserectangle{12} Observe that the proportional relationship of \coloredboxLHH{L-H-L} is not statistically significant within its parent set \coloredboxLHH{H-H-L}. \protect\caserectangle{13} Select \coloredboxLHH{H-H-L} strategy by clicking on \textit{Gene}. \protect\caserectangle{14} Observe the proportional relationships of \coloredboxLLH{L-H-L} and \coloredboxLLH{H-L-L}, sub-strategies to \coloredboxLHH{H-H-L}. \protect\caserectangle{15} Add the formulated strategy to \textit{Operation List} and enter memos. \protect\caserectangle{16} Click \raisebox{-0.4ex}{\includegraphics[height=2ex]{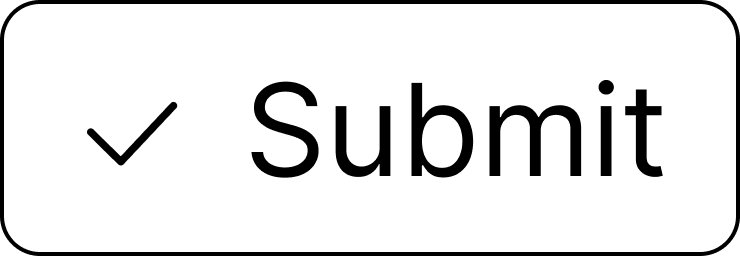}} to retrain the model. \protect\caserectangle{17} Select the retrained model to initiate an iterative exploration. \protect\caserectangle{18} Identify the cluster of three similarly ranked genes, EGFR, MYC, and RPL13, using the \textit{Rank Indicator}. \protect\caserectangle{19} Lasso the three clustered genes.}
  \label{fig:teaser}
}

\graphicspath{{figs/}{figures/}{pictures/}{images/}{./}} 

\usepackage{tabu}                      
\usepackage{booktabs}                  
\usepackage{lipsum}                    
\usepackage{mwe}                       
\usepackage{tabu}                      
\usepackage{booktabs}                  
\usepackage{lipsum}                    
\usepackage{mwe}                       
\usepackage{multirow}
\usepackage{wrapfig}
\usepackage[dvipsnames,svgnames]{xcolor}
\usepackage{times}                     
\usepackage{cite}       
\usepackage{paralist}
\usepackage{balance}

\usepackage{amsthm,amsmath,amssymb}
\usepackage{tikz}

\usepackage{mathrsfs}

\usepackage{mathptmx}                  

\newcommand{\prefix}{\textcolor[RGB]{0, 102, 204}}

\newcommand{\jhr}{}

\definecolor{mycustomcolor}{HTML}{2D2C4C}
\definecolor{casecolor}{HTML}{E5E5E5}

\definecolor{correct}{HTML}{A52A2A}

\newcommand*\caserectangle[1]{\tikz[baseline=(char.base)]{
            \node[rectangle, draw=black, fill=casecolor, text=black, inner sep=2pt] (char) {#1};}}

\newcommand*\darkcircled[1]{\tikz[baseline=(char.base)]{
            \node[shape=circle,draw=mycustomcolor,fill=mycustomcolor,inner sep=0.5pt, line width=0.5pt, text=white, font=\footnotesize] (char) {#1};}}

\usepackage{titlesec}

\newcommand*\circled[1]{\tikz[baseline=(char.base)]{
            \node[shape=circle,draw=black,inner sep=0.5pt, line width=0.5pt, font=\footnotesize] (char) {#1};}}

\newcommand*\specialcircled[2]{%
\tikz[baseline=(char.base)]{
    \node[shape=circle,draw=black,align=center,inner sep=0.5pt, font=\footnotesize, line width=0.5pt] (char) {%
        \scalebox{0.8}{#1}\scalebox{0.6}{#2}};}%
}

\newcommand*\specialcircledg[2]{%
\tikz[baseline=(char.base)]{
    \node[shape=circle,draw=gray,fill=gray,text=white,align=center,inner sep=0.5pt, font=\footnotesize, line width=0.5pt] (char) {%
        \scalebox{0.8}{#1}\scalebox{0.6}{#2}};}%
}

\newcommand*\specialcircledgsingle[1]{%
\tikz[baseline=(char.base)]{
    \node[shape=circle,draw=gray,fill=gray,align=center,text=white,inner sep=0.5pt, font=\footnotesize, line width=0.5pt] (char) {#1};}%
    }

\definecolor{mybgcolorHHH}{HTML}{CAD8FF} 
\definecolor{mybgcolorLHH}{HTML}{CFDFFF} 
\definecolor{mybgcolorLLH}{HTML}{E9F1FF} 
\definecolor{mybgcolorLLL}{HTML}{F0F8FF} 

\definecolor{gene}{HTML}{9ECA80}
\definecolor{cc}{HTML}{FFC488}
\definecolor{bp}{HTML}{98CAF7}
\definecolor{pathway}{HTML}{CDC0DB}
\definecolor{mf}{HTML}{F7A8C3}

\usepackage{tcolorbox}

\newcommand{\coloredboxHHH}[1]{\colorbox{mybgcolorHHH}{\textcolor{black}{\small #1}}}

\newcommand{\coloredboxLHH}[1]{\colorbox{mybgcolorLHH}{\textcolor{black}{\small #1}}}

\newcommand{\coloredboxLLH}[1]{\colorbox{mybgcolorLLH}{\textcolor{black}{\small #1}}}

\newcommand{\coloredboxLLL}[1]{\colorbox{mybgcolorLLL}{\textcolor{black}{\small #1}}}

\begin{document}


\firstsection{Introduction}
\maketitle

\par
\begin{figure}[h]
\centering
\vspace{-3mm}
\includegraphics[width=\linewidth]{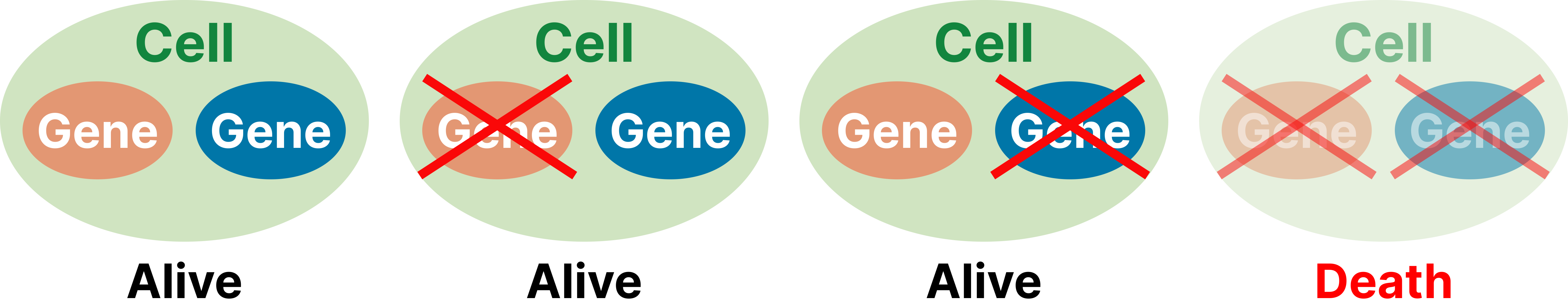}
\vspace{-6mm}
\caption{SL gene pairs refer to a pair of genes where inhibiting one does not impact cancer cell survival, but inhibiting both leads to cell death.}
\label{fig:glyph}
\vspace{-3mm}
\end{figure}

 Synthetic Lethal (SL) relationships among genes denote a specific type of gene interaction characterized by co-expression~\cite{TCSLCAT}. In this context, inhibiting one lethal gene alone does not affect cell survival, but simultaneous inhibition of both genes results in cell death~\cite{gourley2019moving}, as depicted in~\cref{fig:glyph}. This concept is crucial for targeted cancer therapy, as targeting the SL partner gene of a known cancer-related gene allows for precise cancer cell elimination while sparing normal cells. Thus, it offers a viable strategy for inducing the demise of cancer cells when directly targeting certain genes is challenging.

\par Given the substantial medical relevance of SL pairs, screening them is crucial for clinical applications. \jhr{Several experimental biology-based methods, such as RNA interference and genome editing with \textit{CRISPR/Cas9}~\cite{SLC}, have been introduced for comprehensive SL screening~\cite{yang2012genomics}. However, these traditional wet lab methods are resource and labor-intensive, time-consuming, and prone to off-target effects~\cite{SLSTPMC}.} Additionally, validated SL pairs constitute less than 0.1\% of all potential pairs due to the vast number of human gene combinations~\cite{cai2020dual}. To overcome these challenges, computational methods~\cite{Wang2022ComputationalMD}, based on biological insights, have been suggested to speed up and enhance the precision of SL gene pair predictions. These methods span statistical approaches~\cite{feng2019platform}, network-based strategies~\cite{jacunski2015connectivity}, and classic machine learning (ML)~\cite{wu2021synthetic}, along with artificial intelligence (AI) techniques~\cite{luo2020parameterized}. Notably, AI methods have demonstrated superior predictive accuracy. However, aside from achieving accurate predictions, biologists increasingly prioritize understanding the interpretation of SL relationship mechanisms implicit in these AI models. \jhr{Nevertheless, grasping the biological mechanisms underlying the predicted results remains challenging ~\cite{topatana2020advances}. Moreover, the current AI model tools generally lack sufficient interpretability, resulting in a skeptical stance among researchers toward the predictive outcomes in this field.}

\par Prior research~\cite{wang2021kg4sl,zhang2023kr4sl,Wang2023User-Centric,huang2023concept} has aimed to aid biologists in comprehending the behavior of prediction models, facilitating the exploration of mechanisms behind successful predictions. These approaches are often rooted in biological knowledge. For instance, \textit{KG4SL}\cite{wang2021kg4sl} enhanced SL prediction accuracy by directly incorporating various relevant biological factors into the Knowledge Graph (KG) and integrating them with Graph Neural Networks (GNN). Going a step further, \textit{KR4SL}~\cite{zhang2023kr4sl} developed an interpretable AI model for SL prediction. Expanding the scope, efforts have been made to support both domain experts and non-AI-experts in exploring and explaining model behaviors \cite{Wang2023User-Centric,huang2023concept}.




\par Despite aiding target users in understanding parts of model behaviors to enhance interpretability, the methods mentioned above bring new challenges to Human-AI collaboration, which stem from two intertwined dimensions: the symbiotic interplay between \textbf{\textit{biologists' involvement}} and the \textbf{\textit{application of AI}}. On one hand, earlier research has mainly aimed at improving prediction accuracy by incorporating vast amounts of information into the model input~\cite{li2023latent}. However, in efforts to understand SL relationship mechanisms, the complex data used for model input significantly increases the complexity of processing and analysis. Given the intricacies of biological data, some gene or protein interactions might have subtle relevance to SL mechanisms. This subtlety potentially introduces new noise, influencing both prediction outcomes~\cite{zhang2023emerging} and interpretive subgraphs~\cite{PSLHCMGENN}. Hence, a method \textbf{\textit{synergizing with domain expertise becomes crucial to effectively devise reasonable strategies for screening biological knowledge used to train AI models}}. On the other hand, biologists seek to understand AI models and expect clear insights into predictions. This comprehension is vital for decision-making and selecting wet lab directions, allowing experts to trust or work with these models~\cite{jakubik2022empirical}. Within this collaborative process, biologists must carefully \textbf{\textit{consider the division of responsibilities between themselves and the AI models}} while simultaneously evaluating their reliability. In the end, biologists bear the accountability for prediction outcomes. Current approaches mainly provide AI-generated suggestions for biologists to make final decisions~\cite{zhang2023kr4sl}, with little focus on the \textbf{\textit{iterative synergy between AI and biologists}}, such as effective means to rectify the misleading knowledge acquired by the model. Nevertheless, an iterative interaction~\cite{wu2022survey} has the potential to significantly impact the quality and efficiency of the analytical process, which highlights the need for a more integrated and dynamic collaboration between biologists and AI systems.

\par To tackle these challenges, we initiated a substantial partnership with six biologists and SL researchers, immersing ourselves in their daily research routines. Through a formative study involving these professionals, we conducted insightful interviews to discern their needs and expectations. These discussions not only provided an understanding of their requirements, but also yielded six design requirements. These requirements are structured into two tiers: \textbf{\textit{1) Human-Engaged Knowledge Graph Refinement Based on Metapath\footnote{\small{A sequence of node types that guide a walk from the origin to the destination node, commonly applied in biomedical network analysis.}} Strategies}} and \textbf{\textit{2) Cross-Granularity SL Interpretation Enhancement and Mechanism Analysis.}} We propose an iterative framework of human-AI collaboration aligned with these design requirements. Initially, domain experts explore new SL pairs using predictions and interpretive paths\footnote{\small{\jhr{Interpretive paths are explanatory links and nodes connecting the primary gene with the predicted gene through biological relationships.}}} generated by a model trained on the entire data. \jhr{During this process, irrelevant or incorrect paths that may introduce noise to the predictions are eliminated from the KG using appropriate metapath strategies. Subsequently, the model retrains, allowing domain experts to iteratively scrutinize predictions and interpretive paths, refining the KG. This iterative process continuously optimizes predictions and mechanism exploration, enhancing expert participation and intervention, ultimately leading to increased trust. Furthermore, mechanism exploration can identify more persuasive SL pairs for validation through wet lab experiments.} To support this cycle, \textit{SLInterpreter} is developed to help domain experts organize and compare interpretive paths across granularities, uncovering potential SL mechanisms and more persuasive SL pairs. We conducted a case study involving our collaborative experts, followed by interviews to gather their feedback. The findings notably underscored the approach's superiority in delivering informative domain insights, facilitating decision-making, and presenting an \jhr{intuitive} interface. In summary, the contributions of this study can be outlined as follows:
\begin{compactitem}
    \item We collaborated closely with biologists and SL researchers, conducting insightful interviews to uncover six essential design requirements for addressing SL mechanisms.
    \item We introduced the iterative framework \textit{SLInterpreter}, tailored to meet these design prerequisites, empowering experts to explore SL pairs and mechanisms, refine the KG, and foster trust.
    \item We validated the effectiveness of \textit{SLInterpreter} through a comprehensive case study and interviews, demonstrating its superior performance and value.
\end{compactitem}

\vspace{-1mm}

\section{Related work}
\subsection{Computational Methods for Synthetic Lethal Prediction}
\par Constrained by the limitations posed by wet lab experiments in screening SL pairs, numerous computational approaches have been introduced to predict SL pairs~\cite{Wang2022ComputationalMD,yang2021mapping}. These approaches fall into three main categories: \textit{statistical-based}, \textit{network-based}, and \jhr{\textit{ML-based}} methods. 

    
\par \textit{Statistical-based}~\cite{feng2019platform,conde2009human,wang2013identification,deng2019sl} and \textit{network-based} methods~\cite{liu2018synthetic,ku2020integration} model existing SL data under specific assumptions. The interpretability of these methods mainly stems from their ability to reveal statistical or topological patterns associated with validated SL pairs. For instance, \textit{DAISY}~\cite{jerby2014predicting} employed three statistical methods for different cancer sets to infer SL interactions from genomic data. Kranthi et al.~\cite{kranthi2013identification} used a graph-based approach with centrality measures, assuming that highly connected proteins are crucial and their knockout leads to lethality. \jhr{However, these methods may not cover all data aspects, resulting in high prediction accuracy only for the aspects they cover, while accuracy may be low for other data~\cite{Wang2022ComputationalMD}. Additionally, this can introduce biases and challenges in discovering new SL mechanisms.} \jhr{\textit{ML-based} methods are designed to identify intricate patterns in large datasets that defy straightforward analysis, with fewer and more relaxed assumptions about the data~\cite{ij2018statistics}, thus improving prediction accuracy for unobserved data.} For instance, Wu et al.~\cite{wu2021synthetic} developed a k-NN model, suggesting that gene pairs similar to validated SL pairs have higher SL potential. Lundberg S.M.~\cite{lundberg2020local} enhanced tree-based models' interpretability by measuring local feature interactions and using these insights to understand the model's global structure. However, most of these classic ML methods still depend on the manual extraction of gene features based on past knowledge, which can introduce human biases\jhr{~\cite{Wang2022ComputationalMD}}. While deep learning methods reduce the need for manual feature engineering, they often lack transparency and function as opaque black boxes.
    

\par Given the limitations of the aforementioned methods, the KG-based GNN model leverages the extensive validated facts within the KG to learn gene representations and predict SL relationships with minimal or no explicit feature engineering~\cite{velivckovic2017graph}. \jhr{The KG-based GNN model has demonstrated state-of-the-art predictive accuracy and provides interpretive paths with their weights through attentive aggregation. Consequently, we built our backend based on the KR4SL framework~\cite{zhang2023kr4sl}.}

\vspace{-1mm}

\subsection{Graph Neural Network Interpretability}
\par Interpreting GNN models is crucial, prompting many efforts to enhance their interpretability~\cite{luo2020parameterized,shan2021reinforcement,spinelli2022meta,ying2019gnnexplainer,yu2023improved}. These efforts are primarily categorized into three levels of granularity~\cite{zhang2023mixupexplainer}: \textit{1) Instance-Level Explanation}: This aims to elucidate the prediction for each instance by identifying significant substructures or subgraphs. \textit{2) Model-Level Explanation}: This focuses on comprehending the global decision rules captured by the GNN. \textit{3) Group-Level Explanation}: This provides explanations for a group of predicted instances.

\par Most GNN interpretability methods focus on \textit{Model-Level Explanation}~\cite{baldassarre2019explainability,luo2020parameterized,spinelli2022meta}, aimed at helping data or AI experts understand the structures and behaviors of models or assist in debugging. Nevertheless, these methods often overlook the challenges faced by non-AI domain experts. To address this, some studies have sought to create more intuitive explanations for non-AI experts using \textit{Instance-Level Explanation}~\cite{shan2021reinforcement,yuan2021explainability}. For instance, \textit{GNNExplainer}~\cite{ying2019gnnexplainer} identifies critical subgraphs to elucidate GNN model predictions.


\par While \textit{Instance-Level Explanation} methods enhance domain experts' understanding of a model's decisions for specific instances, they fall short for those concerned with potential mechanisms indicated by the model. To address this limitation, \textit{DrugExplorer}~\cite{Wang2023User-Centric} introduces \textit{Group-Level Explanation}, which allows users to categorize drugs based on similar action mechanisms. Inspired by their approach, our study assists domain users in exploring potential mechanisms at the \textit{Group-Level} by analyzing relationships between instances within interpretive paths throughout iterative interactions with AI. \jhr{Notably, our target users are biologists researching SL. Unlike \textit{DrugExplorer}, which focuses on drug-to-disease relationships, we emphasize gene-to-gene relationships. Furthermore, we enable users to iteratively refine AI predictions and discover new SL mechanisms through cross-granularity exploration.}

\par Additionally, in GNN research, \jhr{\textit{CorGIE}~\cite{liu2022corgie} explores graph topology, node features, and embeddings interplay to decipher GNN operations. Building on this, \textit{GNNLens}~\cite{jin2022gnnlens} provides a visual tool assisting ML experts in identifying potential error patterns, offering insights for optimizing model structures. However, these tools are unsuitable for non-AI domain experts to explore predictions and underlying mechanisms.} Our work differs by focusing on assisting \textbf{domain experts} in a deeper exploration of the underlying mechanisms suggested by GNN models. Through iterative and cross-granularity visualization, we enable domain experts to rectify knowledge erroneously acquired by GNN models, facilitating a more profound understanding of predictions.

\vspace{-1mm}
\subsection{Human-AI Collaboration}

\par \jhr{\textit{Human-AI collaboration} refers to the interactive cooperation between humans and AI to jointly address specific tasks, with responsibilities divided based on the strengths of each party~\cite{mackeprang2019discovering, shi2023RetroLens}. As AI performance continuously improves, the ways humans and AI collaborate are constantly evolving, with three main patterns proposed~\cite{shi2023RetroLens}. The first pattern is \textit{AI-assisted decision-making}, where AI provides necessary information and support for humans to make final decisions. Many studies have seamlessly integrated AI assistance into the workflow of domain experts~\cite{oakes2023workflow, wang2022Explanations,zhang2020AI-Assisted}. The second pattern, \textit{Human-in-the-loop}, involves incorporating expert intervention into the process of training AI models. This approach aims to train accurate models at minimal cost by integrating human knowledge and experience~\cite{wu2022survey,dudley2018review,ma2023modeling,zhao2022Human-in-the-loop,mackeprang2019discovering}. The third pattern, \textit{joint action}, involves humans and AI working together towards a common objective, functioning as a single entity~\cite{zuckerman2022tangible,ashktorab2020human,lai2020chicago,lai2019human,lai2022human,mackeprang2019discovering}.}

\par \jhr{Building upon existing research, our approach seamlessly integrates the expertise of biologists with the supportive role of AI. Biologists first explore new potential SL pairs through the predictions and interpretive paths generated by the model trained on entire data. Subsequently, we facilitate biologists' active involvement in a \textit{human-in-the-loop} process, assisting them in refining the knowledge acquired by the model through appropriate metapath strategies, which results in interpretive paths that progressively reflect biologists' expertise and biological principles.}

\vspace{-1mm}
\section{Formative Study}
\par The formative study aimed to thoroughly understand and evaluate the challenges faced by SL researchers, their preferences and expectations regarding collaboration with AI, and their system design requirements. To achieve this, we conducted semi-structured interviews with four researchers (\textbf{E1-E4}) specializing in SL mechanisms and two clinical physicians (\textbf{E5-E6}) from a local university and a hospital (Mean Age = 47.5, SD = 46.6, four males and two females). Within this group, \textbf{E1-E2} specialize in screening SL pairs using the \textit{CRISPR/Cas9} technique, focusing on thyroid cancer and glioma, respectively. \textbf{E3-E4} are engaged in anticancer drug research, collaborating with \textbf{E5-E6}. Each participant has extensive research experience in SL and possesses knowledge or experience in AI-assisted prediction of SL pairs. Through thematic analysis~\cite{clarke2017thematic}, we extracted valuable insights regarding the challenges faced by domain experts and summarized the design requirements for our approach. Each interview session lasted approximately $45$ minutes.

\vspace{-1mm}

\subsection{Experts' Conventional Practices}
\par Our experts emphasized their reliance on wet-lab experiments or statistical methods in the past for discovering and validating new SL pairs. Specifically, wet lab experiments require skilled researchers to culture relevant cells. Moreover, specific gene inhibitors are utilized, necessitating manual observation and recording of experimental data. For instance, in tumor experiments, it is crucial to observe and record proliferation-related experimental data. Some tumors require the assessment of their migration properties through methods like cell scratching, followed by cell washing or drug addition observation. If the experiment progresses smoothly, it typically involves about five researchers and takes approximately half a year. Considering potential issues such as unexpected cell death or microbial contamination during cell culture, the duration may extend further. Therefore, given the considerations of experimental costs, manpower, and equipment needs, most labs are limited to conducting one to two wet lab experiments at a time.

\par Given the high cost and limitations of wet lab experiments, our experts emphasized their need for a method that can predict SL pairs at a lower cost and with higher accuracy, while also providing reasonable explanations. To this end, they have attempted to use AI-assisted methods for experimental pair screening. However, in their attempts, they found that current AI methods lack sufficient interpretability and tools for mechanism exploration. Moreover, since the models can only provide one-way prediction results, applying their expertise to intervene in the AI prediction process proved to be difficult. The sole form of interaction experts have with the model is to accept or reject its predictions, which leads to a lack of trust in these models' predictions.



\vspace{-1mm}

\subsection{Experts' Concerns and Barriers}

\par Domain experts initially voiced concerns over biological data accuracy. Several interviewees highlighted the importance of reliable and pertinent data in training AI models. They underscored that ``\textit{...this basically decides how accurate the AI predictions are and how sensible the explanations it gives make sense.}'' For instance, \textbf{E1} stated, ``\textit{Picking the right data to train the model is really important, and where that data comes from is definitely something to seriously think about.}'' \textbf{E2} mentioned, ``\textit{A worry we all share is that if the model is learning, the info we feed it better is true... When we're looking into things, if we spot any issues with the data or conclusions, I hope there's a way to quickly filter them out.}'' This reveals the first challenge: \textbf{C1. Reliability of biological data in AI training}. As experts analyze the model's interpretive paths, having effective methods and interactive tools for screening data and addressing potential errors or misleading conclusions is vital.

     
\par In terms of collaboration with AI, several interviewees expressed concerns about the current pattern. The prevalent human-AI collaboration approach in SL involves domain experts passively accepting prediction results. Subsequently, experts simply choose a few gene pairs for initial experimental validation, typically those not yet investigated or related to their current research direction. This collaboration mode entirely overlooks the active involvement of domain experts in the prediction task, leading to researchers distrusting the conclusions. As mentioned by \textbf{E4}, ``\textit{I simply get the results the model spits out, and when I see some obvious mistakes, there's no way for me to fix [them]. It just keeps eroding my trust in AI.}'' This highlights the second challenge: \textbf{C2. Lack of intervention in AI prediction process}.

\par Interviewees also expressed concerns in decision-making, potential SL mechanism exploration, and continuous collaboration with AI. Specifically, determining whether to accept and validate AI's prediction results through experiments poses a significant challenge. As highlighted by \textbf{E3}, ``\textit{We're talking about over $20,000$ human genes here, and doing experiments for all those gene pairs? That's just not economically achievable.}'' Similarly, \textbf{E2} pointed out, ``\textit{...So, AI may predict a bunch of gene pairs, but it takes us at least half a year to check out a few pairs at the same time. So, we really hope the predictions are highly accurate and AI can give us a good explanation of how they came up [with the predictions].}'' This reveals the third challenge: \textbf{C3. High trial-and-error costs}. Wet lab experiments are time-consuming and resource-intensive. Researchers cannot experimentally validate all predictions, and careful consideration is needed in selecting the next research targets. For AI predictions, there is a need for more abundant and reasonable evidence to demonstrate their experimental value.

\par Furthermore, even if the model provides a plausible explanation for individual gene pairs, domain experts express greater concern about extracting underlying mechanisms and identifying commonalities from multiple prediction results. For example, \textbf{E2} pointed out, ``\textit{The prediction results contain too many possible interpretive paths. Although these paths are sorted by scores, we still need to focus on at least 10 or so paths, and once too many entities are set up, it puts a very heavy burden on us to explore possible mechanisms}'', highlighting the challenge: \textbf{C4. Limitations in SL mechanism exploring}. While domain experts find the commonalities among these explanations intriguing, current methods remain cumbersome and insufficient in assisting experts to effectively compare and analyze these commonalities.

\par While some SL experts acknowledge the accuracy and efficiency of AI predictions, researchers still harbor widespread concerns regarding sustaining ongoing collaboration with AI. \textbf{E5} noted, ``\textit{We've got some AI-related projects going on, but they're mostly just starting out... Honestly, AI models are still kind of new to us compared to the usual methods, so we stick to what we know best - traditional research methods using stats or genomics.}'' Similarly, \textbf{E6} commented, ``\textit{Yeah, I totally agree that we need [tools] to fix AI when it learns wrong stuff. In the past, we didn't have such support, and we're not used to working with interactive models like that... So, if those tools come along, we've gotta make sure they actually work by checking how they affect the model's performance,}'' highlighting the final challenge: \textbf{C5. Lack of experience in iterative collaboration with AI}.

\vspace{-1mm}

\subsection{Experts' Needs and Expectations}
\par After conducting interviews with experts, we've compiled a set of requirements aimed at effectively addressing the discussed challenges. Our approach is tailored to seamlessly integrate with conventional practices, allowing experts to grasp prediction model behaviors and delve into the mechanisms behind successful SL predictions. To delve deeper, we systematically tackle these requirements in two parts: {\prefix{\texttt{\textbf{[KG Refinement]}}}}\textit{Human-Engaged Knowledge Graph Refinement Based on Metapath Strategies} and {\prefix{\texttt{\textbf{[SL Interpretation]}}}}\textit{Cross-Granularity SL Interpretation Enhancement and Mechanism Analysis}. These parts operate in an {\prefix{\texttt{\textbf{[Iterative Cycle]}}}}, continuously refining model predictions and exploring mechanisms, thereby enhancing expert involvement and intervention, and bolstering trust.


\par {\prefix{\texttt{\textbf{[KG Refinement]}}}}Initially, researchers underscored the urgency of conducting a rapid and comprehensive exploration of training data and interpretive paths, which is essential for gaining understanding and familiarity with both the model's input data and its performance. To tackle challenges concerning the reliability of biological data \textbf{[C1]}, providing a concise summary of training data and interpretive paths is essential. This forms the basis for our first design requirement: \textbf{DR1. Offer a clear summary of training data, prediction results, and interpretive paths}. Furthermore, addressing the urgent need expressed by several researchers to quickly identify erroneous or noisy data in the training set through the error-prone interpretive paths indicated by the model, our second design requirement emerges: \jhr{\textbf{DR2. Detect relevant training data corresponding to erroneous or noisy paths}}.



\par {\prefix{\texttt{\textbf{[KG Refinement]}}}}Addressing the second challenge \textbf{[C2]}, which concerns researchers' skepticism towards AI predictions due to inadequate intervention, we introduce \textbf{DR3. Provide appropriate interventions to filter data corresponding to interpretive paths}. \jhr{This requirement was consistently emphasized during semi-structured interviews, highlighting researchers' worries about the negative impact of insufficient intervention on AI predictions.} To alleviate these concerns, our approach offers a comprehensive visual analysis of potentially erroneous or noisy data, along with interactive features allowing users to correct any identified issues promptly.

\par {\prefix{\texttt{\textbf{[SL Interpretation]}}}}To address the third challenge \textbf{[C3]}, it is proposed that the system should not only enhance the accuracy of predicting gene pairs through proper data input but also offer a clear and sufficient representation of interpretive paths, considering the high cost of trial and error. This entails the ability to vividly present information about both the genes and selected entities, along with the attributes of edge relationships. The emphasis lies on minimizing visual clutter and ensuring the smooth presentation of both topological and textual information. During semi-structured interviews, \textbf{E1} stressed that ``\textit{only when the interpretive paths are presented clearly and intuitively, and align well with our domain expertise, will the predicted results be convincingly justified for wet lab experiments validation.}'' This underscores the importance of our fourth design requirement: \textbf{DR4. Provide detailed displays of paths and entities for interpretive paths.}


 \par {\prefix{\texttt{\textbf{[SL Interpretation]}}}}Regarding the fourth challenge, identified by researchers in identifying common potential mechanisms from AI predictions \textbf{[C4]}, the system must clearly present and analyze similarities across various interpretive paths, which is crucial as it can significantly aid researchers in comprehending existing mechanisms and uncovering new ones. Acknowledging the complexity researchers face in exploring common potential mechanisms from numerous paths, the design must balance cognitive load with information presentation. According to \textbf{E1}, ``\textit{Exploring potential mechanisms in SL is highly complex, and current AI struggles to directly accomplish theoretical derivations.}'' Additionally, as noted by \textbf{E2}, ``\textit{However, identifying commonalities in AI predictions can suggest fruitful directions, provided these commonalities are easily extracted and summarized. Comparing them individually in \jhr{common AI-assisted systems with limited attention is impractical and can narrow our focus.}}'' This leads us to our fifth design requirement: \textbf{DR5. Comparative analysis and mechanism exploration of different prediction results and interpretive paths.}

\par {\prefix{\texttt{\textbf{[Iterative Cycle]}}}}In response to \textbf{[C5]}, we introduce the last design requirement: \textbf{DR6. Facilitating iterative Human-AI interaction with feedback and recordings}. This requirement is crafted to support domain experts in overcoming unfamiliarity and uncertainty during their collaboration with AI systems. By ensuring a seamless exchange of feedback and recording interactions, we empower domain experts to grasp the repercussions of their interventions on both datasets and model performance through ongoing engagement. The essence of this requirement lies in capturing users' actions comprehensively, thus enabling them to retrace their steps when faced with unsatisfactory intervention outcomes or when model metrics fall short of expectations. This iterative process fosters domain experts' familiarity with AI collaboration, allowing them to refine model performance progressively. Echoing the sentiments of \textbf{E5}, ``\textit{During the early stages of AI interaction, keeping up with iterative exploration is key. You know, being able to jot down all those steps you're taking? It's like having a handy tool that lets you go back and forth, compare stuff, and double-check what you've done. This way, you can make sure your interventions are on point and your analysis is as sharp as it can be.}''


\section{SLInterpreter}

\begin{figure}[h]
\centering
\vspace{-4mm}
\includegraphics[width=\linewidth]{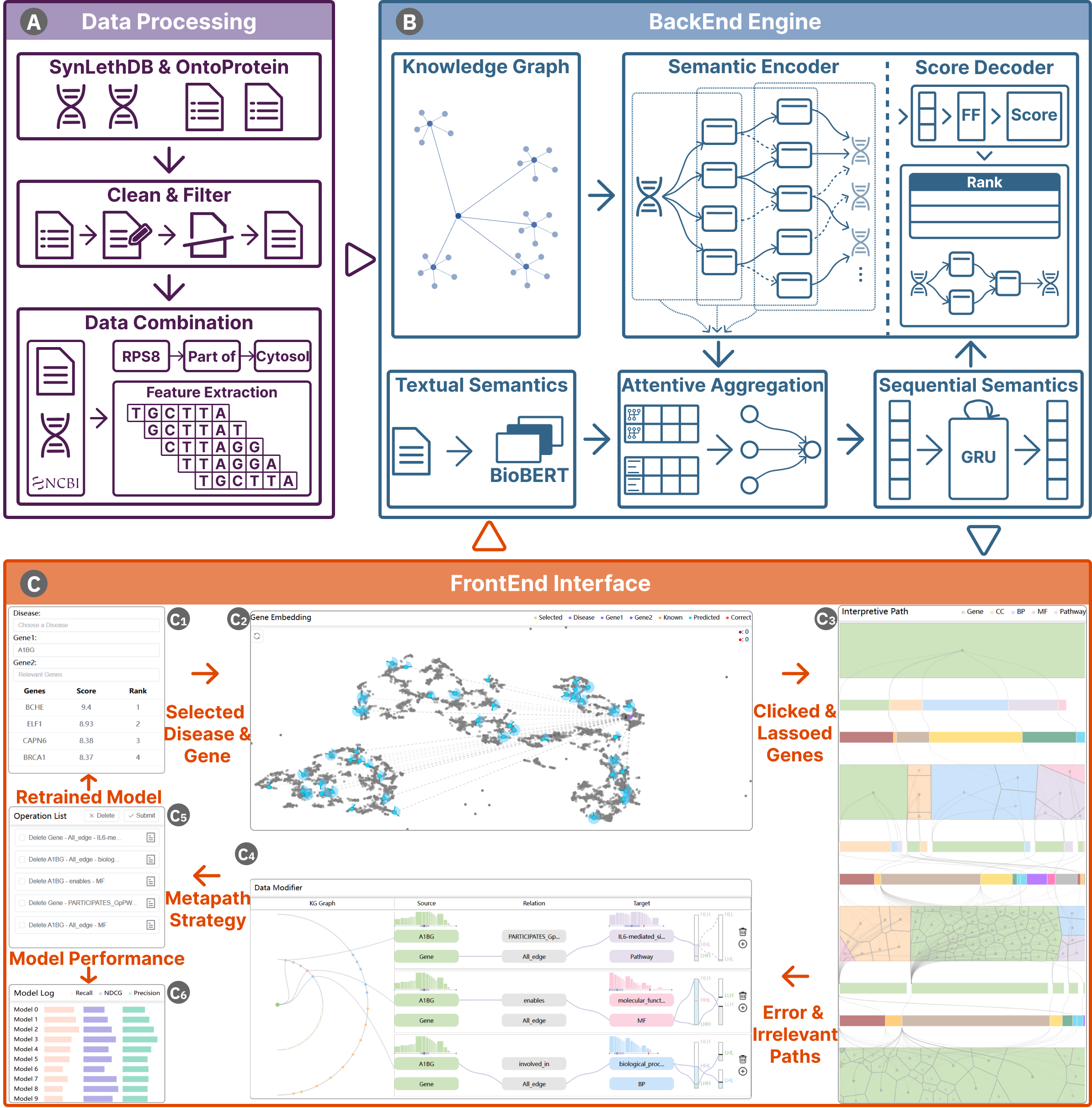}
\vspace{-6mm}
\caption{\textit{SLInterpreter} comprises \protect\specialcircledgsingle{A} data processing module, \protect\specialcircledgsingle{B} backend engine, and \protect\specialcircledgsingle{C} frontend interface, facilitating an iterative workflow.}
\label{fig:pipeline}
\vspace{-3mm}
\end{figure}

\par Our analysis approach, implemented through \textit{SLInterpreter}, consists of three main components: \specialcircledgsingle{A} data processing module, \specialcircledgsingle{B} backend engine, and \specialcircledgsingle{C} frontend interface (\cref{fig:pipeline}). The data processing module prepares data for the KG and extracts gene features. The backend engine trains the GNN model with the KG and gene features, predicting SL pairs and interpretive paths. The frontend then provides multiple interconnected views for iterative and cross-granularity SL analysis.

\vspace{-1mm}

\subsection{Data Description and Processing}
\par We provide detailed information regarding the data utilized in the backend engine, including both \textit{Gene Feature} and \textit{Knowledge Graph}.

\par \textbf{Gene Feature.} The gene feature data includes both gene sequence $\mathbb{S}$ and gene description $\mathbb{D}$ sourced from the National Center for Biotechnology Information (NCBI) database \cite{geer2010ncbi}, recognized as an authoritative gene database in the biological domain, ensuring data reliability and professionalism. Each gene sequence consists of approximately $19,000$ base pairs (ACGT). However, directly converting these sequences into numerical form and performing operations like dimensionality reduction would entail significant computational costs due to the extensive number of genes and the relatively lengthy gene sequences for each. To address this challenge, we utilize the \textit{k-mer} method~\cite{ghandi2014kmer}
, a common method of DNA feature extraction in the biological field, to identify local patterns in gene sequences $\mathbb{S}$. Meanwhile, we utilize \textit{BioBERT}~\cite{lee2020biobert} to process functional and descriptive gene information $\mathbb{D}$.

\par \textbf{Knowledge Graph.} The biological data used to build the \jhr{Knowledge Graph (KG)} is sourced from two main repositories (\cref{tab:table_data_description}): the SL knowledge graph \textit{SynLethDB}~\cite{wang2021kg4sl} and the biomedical knowledge dataset \jhr{\textit{ProteinKG25}~\cite{zhang2022ontoprotein}}. These datasets include validated gene SL interactions and external knowledge about gene functions, such as pathways and biological processes. The resulting KG comprises five types of entities: \textit{Gene}, \textit{Pathway (PW)}, and three types of \textit{Gene Ontology (GO)}: \textit{Biological Process (BP)}, \textit{Molecular Function (MF)}, and \textit{Cellular Component (CC)}. Additionally, it incorporates multiple relationship types, primarily indicating gene associations with specific paths or annotations by \textit{GO}. \jhr{Entities are connected by these relationships to form triplets, which are the fundamental components of KG structures, expressing how two entities are linked (e.g., \textit{RPS8}---\textit{Part of}---\textit{Cytosol}). Collectively, these triplets constitute the final KG.}

\begin{table}[h]
\vspace{-4mm}
\caption{Data Description.}
\label{tab:table_data_description}
\centering
\vspace{-3mm}
{\small
\begin{tabu}{cccc}
\toprule
    & \textbf{\#Entities}   & \textbf{\#Relations}  & \textbf{\#Triples}\\
\midrule
SL Graph& 9746  & 1  & 35374\\
\jhr{ProteinKG25} & 42547 &  32 & 361245\\
\textbf{Knowledge Graph}& 42547 & 33 & 396619\\
\bottomrule
\end{tabu}}
\vspace{-6mm}
\end{table}

\subsection{Backend Engine}
\par Predicting SL gene pairs can be seen as akin to link prediction. Following the methodology described in \textit{KR4SL}~\cite{zhang2023kr4sl}, we employ a GNN-based model. This model predicts potential SL relationships among gene pairs lacking direct connections by utilizing established relationships to generate prediction paths and allocate weights to these paths, reflecting the strength of the predicted SL relationship. 

\par \textbf{Training Data Construction.} According to \textit{KR4SL}~\cite{zhang2023kr4sl}, we begin by constructing a directed heterogeneous KG by integrating the well-established SL graph \textit{SynLethDB}~\cite{wang2021kg4sl} with the comprehensive biological dataset \textit{ProteinKG25}~\cite{zhang2022ontoprotein}. \jhr{Notably, merging these datasets does not increase the number of entities since all human genes in \textit{SynLethDB} are already included in \textit{ProteinKG25}.} This augmented graph serves as the input for the GNN model, enabling it to capture a rich set of relational patterns and interactions among genes. \jhr{For transparency, we have cited the public databases used in this study within our open-source project\footnote{\small{\url{https://github.com/jianghr-shanghaitech/SLInterpreter-Demo}}}.} \jhr{Additionally, in the context of \textit{SLInterpreter}, paths that are irrelevant or less related to SL are iteratively pruned from the KG, refining the input for the model.}


\par \textbf{Predicting SL Pairs.} Following the encoder-decoder architecture in \textit{KR4SL}, we utilize the heterogeneous KG structure to identify potential SL pairs by tracing relational paths. This approach integrates structural semantics from the graph with textual semantics extracted from gene descriptions. Additionally, it employs attentive aggregation among triples with the same target, enhancing semantic flow through \textit{GRU modules} at each layer and accounting for the paths' weights. In the final encoder layer, candidate SL partners are pinpointed and evaluated by a \textit{score decoder}. The decoder ranks candidates by their potential to form an SL relationship with the target gene, selecting the top-50 as predicted SL partners, \jhr{a commonly used metric in such link prediction tasks~\cite{zhang2023kr4sl}}. Simultaneously, the model outputs the \jhr{3-hop} path containing the predicted link as an interpretive path along with its score. 


\par \jhr{The triplets in the dataset are divided into 80\% training, 10\% validation, and 10\% test sets.} Performance evaluation focused on precision for the top-50 candidates in the test set. The experiments, detailed in Table 1 in the Appendix, highlighted the model's efficacy, achieving a precision rate of 53\% for the top-50 candidates. In comparison, \jhr{using the same data}, the average precision of other similar models~\cite{wang2021kg4sl,zhu2023slgnn,wang2022NSF4SL} was 16.29\%, with the second-best model, \textit{NSF4SL}~\cite{wang2022NSF4SL}, achieving 34.7\%, which demonstrates the effectiveness of this model in detecting key gene interactions, providing valuable insights for cancer research.

\vspace{-1mm}

\subsection{Frontend Interface}
\par In collaboration with domain experts, we developed a frontend interface for exploring biological and genetic data for SL analysis, featuring: 1) \textit{Session View} (\prefix{\texttt{\textbf{[KG Refinement]}}}): Includes disease and gene search boxes, a log of modifier operations, and model performance overview. 2) \textit{Embedding View} (\prefix{\texttt{\textbf{[SL Analysis]}}}): Provides an overview of prediction results on feature reduction. 3) \textit{Interpretation View} (\prefix{\texttt{\textbf{[SL Analysis]}}}): Displays cross-granularity interpretation paths, and 4) \textit{Modifier View} (\prefix{\texttt{\textbf{[KG Refinement]}}}): Shows KG representations for gene and modifier selection in metapath strategies. After setting the metapath strategies, users can start a new exploration cycle based on a retrained model ({\prefix{\texttt{\textbf{[Iterative Cycle]}}}}).

\vspace{-1mm}

\subsubsection{Session View}
\par The \textit{Session View} (\cref{fig:pipeline}-\specialcircledg{C}{1} \& \specialcircledg{C}{5} \& \specialcircledg{C}{6}) offers a platform to interact with the model's prediction outcomes and to review previous logs and model performance. This view comprises three subviews: the \textit{Panel}, the \textit{Operation List}, and the \textit{Model Log}.

\par \textbf{Panel.} Within the \textit{Panel} (\cref{fig:pipeline}-\specialcircledg{C}{1}), three search boxes are designed for selecting prediction results at varying levels. The \textit{Disease search box} (\cref{fig:teaser}-\caserectangle{1}) showcases genes linked to a specific disease, while the \textit{Primary Gene search box} (\cref{fig:teaser}-\caserectangle{3}) enables direct identification of genes of interest through auto-complete. Upon selecting a \textit{Primary Gene}, \textit{Partner Gene Table} (\cref{fig:teaser}-\caserectangle{4}) below the search boxes displays the top-50 \textit{Partner Genes} from the prediction results in ascending order, detailing \textit{Name}, \textit{Score}, and \textit{Rank}. Correctly predicted \textit{Partner Genes} are highlighted in \textcolor{correct}{brown} font. \jhr{We verify the accuracy of predicted \textit{Partner Genes} using ground truth data from \textit{SynLethDB}, sourced from previous wet lab experiments or relevant literature~\cite{wang2021kg4sl}.} Moreover, users can engage with the table by selecting gene rows, which generates tags in the \textit{Partner Gene search box} (\cref{fig:teaser}-\caserectangle{5}). Concurrently, the \textit{Embedding View} and \textit{Interpretation View} update all prediction results related to the selected \textit{Partner Gene}. \textbf{Operation List.} The \textit{Operation List} (\cref{fig:pipeline}-\specialcircledg{C}{5}) records user interactions with the KG in chronological order, featuring collapsible text boxes beneath each action for users to elaborate on their operations and corresponding notes (\textbf{DR6}). Users can select metapaths using checkboxes. Undesired metapaths can be removed by clicking \raisebox{-0.4ex}{\includegraphics[height=2ex]{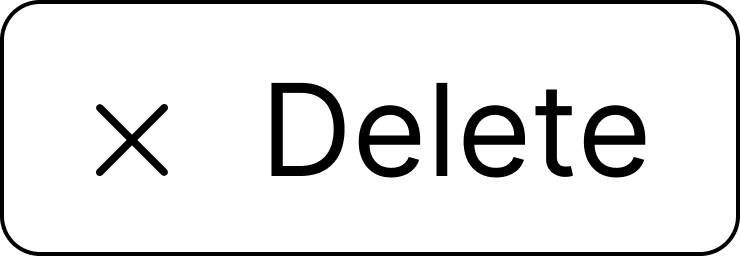}}. Once the selection of metapaths is finalized and \raisebox{-0.4ex}{\includegraphics[height=2ex]{figs/Group-496.png}} is clicked, The backend automatically deletes the corresponding paths, retrains the model, and displays its performance in the \textit{Model Log}. \textbf{Model Log.} The \textit{Model Log} (\cref{fig:pipeline}-\specialcircledg{C}{6}) displays \jhr{grouped bars} detailing model performance indicators, including \textit{Recall}, \textit{NDCG}, and \textit{Precision}. Indicators are color-coded with bar length representing values. The vertical arrangement allows intuitive performance comparison across models. Users can access models by clicking the corresponding bars.

\vspace{-1mm}

\subsubsection{Embedding View}
\par The \textit{Embedding View} (\cref{fig:pipeline}-\specialcircledg{C}{2}) offers users an overview of clustering based on gene features, along with model prediction results and validated SL partners (\textbf{DR1}), enabling exploration of SL pairing patterns within comparable gene groups (\textbf{DR5}). To fully leverage gene attributes, including sequences $\mathbb{S}$ and descriptions $\mathbb{D}$, we use concatenation fusion methods to merge the processed data. We then apply \textit{UMAP}~\cite{McInnes2018UMAPUM} to generate the embedding result due to its efficiency in compressing high-dimensional data into a lower-dimensional representation. Our experiments (Appendix Fig. 1\&2\&3) showed that \textit{UMAP} outperformed \textit{PCA}~\cite{wold1987principal} and \textit{t-SNE}~\cite{Maaten2008VisualizingDU} in preserving global data structure and execution speed. \textit{PCA} often overlooks crucial biological variations~\cite{1339264}, while \textit{t-SNE} primarily reveals local structures without effectively showing similarities between clusters~\cite{linderman2019clustering}. Both \textit{PCA} and \textit{t-SNE} also perform slowly with large datasets. \jhr{Although \textit{UMAP} is non-linear, meaning distances in the \textit{Embedding View} are not linear, it effectively preserves local distances over global distances, enhancing clustering within local regions, which makes it suitable for our focus on the correspondence between local clusters. Consequently, the dashed lines connecting nodes illustrate pairing relationships between genes rather than distance comparisons, helping users identify patterns where genes from one cluster frequently link to genes from another cluster.}

\par
\begin{figure}[h]
\centering
\vspace{-3mm}
\includegraphics[width=\linewidth]{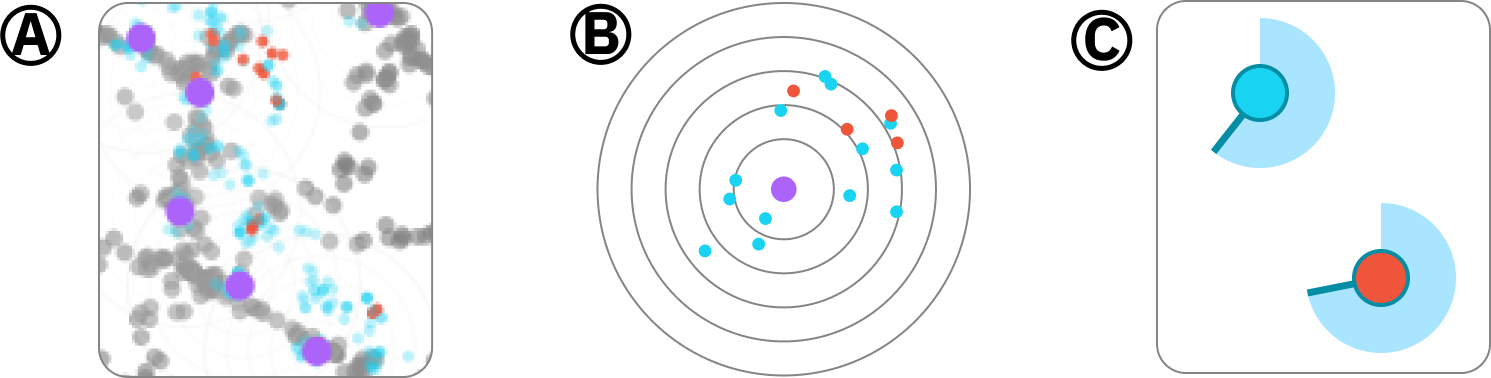}
\vspace{-6mm}
\caption{Glyph Designs for \textit{Embedding View}. \protect\circled{A} is the overview of \textit{Radar} in disease searching. \protect\circled{B} is the enlarged \textit{Radar}. \protect\circled{C} is the \textit{Rank Indicator}.}
\label{fig:embedding}
\vspace{-3mm}
\end{figure}

 \textbf{Visual Design.} The \textit{Embedding View} enhances disease exploration through intuitive visual cues. Upon selecting a disease from the \textit{Disease search box}, the \textit{Embedding View} highlights associated genes with distinct \textcolor[RGB]{171, 99, 250}{\Large{$\bullet$}} points. \jhr{The surrounding \textit{Radars} directly project the relative locations of \textit{Partner Genes} onto concentric circles centered on the \textit{Primary Gene}}. The top-50 predicted \textit{Partner Genes} are highlighted in \textcolor[RGB]{25, 211, 243}{\Large{$\bullet$}} and correctly predicted \textit{Partner Genes} are shown in \textcolor[RGB]{239, 85, 59}{\Large{$\bullet$}} (\cref{fig:embedding}-\circled{A}). Users can employ semantic zooming to manage overlapping points or observe radar patterns more clearly. Hovering over a highlighted point reveals an enlarged \textit{Radar} (\cref{fig:embedding}-\circled{B}), facilitating the identification of tightly clustered \textit{Partner Genes} with common pairing patterns.

\par Once a \textit{Primary Gene} is identified, users can select it in the \textit{Primary Gene search box} for further investigation. Upon selecting, the corresponding \textit{Primary Gene} remains highlighted, while other genes revert to the default color. Simultaneously, the \textit{Embedding View} highlights top-50 predicted \textit{Partner Genes} \textcolor[RGB]{25, 211, 243}{\Large{$\bullet$}} with dashed lines connecting to the \textit{Primary Gene}, validated \textit{Partner Genes} of the \textit{Primary Gene} \textcolor[RGB]{255, 161, 90}{\Large{$\bullet$}}, and correctly predicted \textit{Partner Genes} \textcolor[RGB]{239, 85, 59}{\Large{$\bullet$}}. Each top-50 predicted \textit{Partner Gene} is encircled by \textit{Rank Indicator}  (\cref{fig:embedding}-\circled{C}), an arc indicating its rank of the \textit{Partner Gene}. The arc starts at the top and lengthens clockwise as the rank increases, with a bar representing the end of the arc, helping to discover clusters of similar ranked genes at a glance. When tags are created in the \textit{Partner search box}, corresponding points are highlighted \textcolor[RGB]{99, 110, 250}{\Large{$\bullet$}}, and the dashed line turns solid.

\par \textbf{Interaction.} When no \textit{Disease} or \textit{Primary Gene} is selected, users can lasso all genes to explore patterns among genes with similar features, with the selected genes highlighted \textcolor[RGB]{182, 232, 128}{\Large{$\bullet$}}. Once a \textit{Disease} is chosen, lassoing among disease-related genes helps explore patterns within genes linked to the same disease. Similarly, when a \textit{Primary Gene} is selected, lassoing among the top-50 \textit{Partner Genes} enables the exploration among \textit{Partner Genes} belonging to the same \textit{Primary Gene}.

\vspace{-1mm}

\subsubsection{Interpretation View}
\par The \textit{Interpretation View} (\cref{fig:pipeline}-\specialcircledg{C}{3}) offers users an across-granularity analysis of paths: spanning from comprehensive cross-gene analyses (\textbf{DR5}) to intricate single-path analyses (\textbf{DR4}). To enhance the clarity of metapath content and flow, we break them down based on path hierarchy into three primary elements: 1) the name, type, and proportion of the current hierarchical node; 2) the flow from the current hierarchical node type to the category of the subsequent hierarchical node, and 3) the names of the paths linking to the next hierarchical node.


\par \textbf{Visual Design}\textbf{.} To effectively illustrate the name, type, and proportion of entities on each layer while mitigating excessive overlapping of connecting lines, we employ Voronoi Treemaps~\cite{balzer2005Voronoi}. This method uses the weight of different entities within interpretive paths as their value, dividing space into irregular polygons, optimizing display space utilization, and offering clear differentiation between various entities. \jhr{Clicking on an entity's polygon will highlight other entities connected to it.} Consequently, \jhr{entities belonging to the same category, such as genes or biological processes (BP), }are assigned the same color and grouped together within the same area. \jhr{There are a total of five categories of entities: \textcolor{gene}{Gene}, \textcolor{cc}{Cellular Component (CC)}, \textcolor{bp}{Biological Process (BP)}, \textcolor{mf}{Molecular Function (MF)}, and \textcolor{pathway}{Pathway}, as well as 33 types of path relationships, which are represented using consistent colors.}


\par
\begin{figure}[h]
\centering
\vspace{-3mm}
\includegraphics[width=\linewidth]{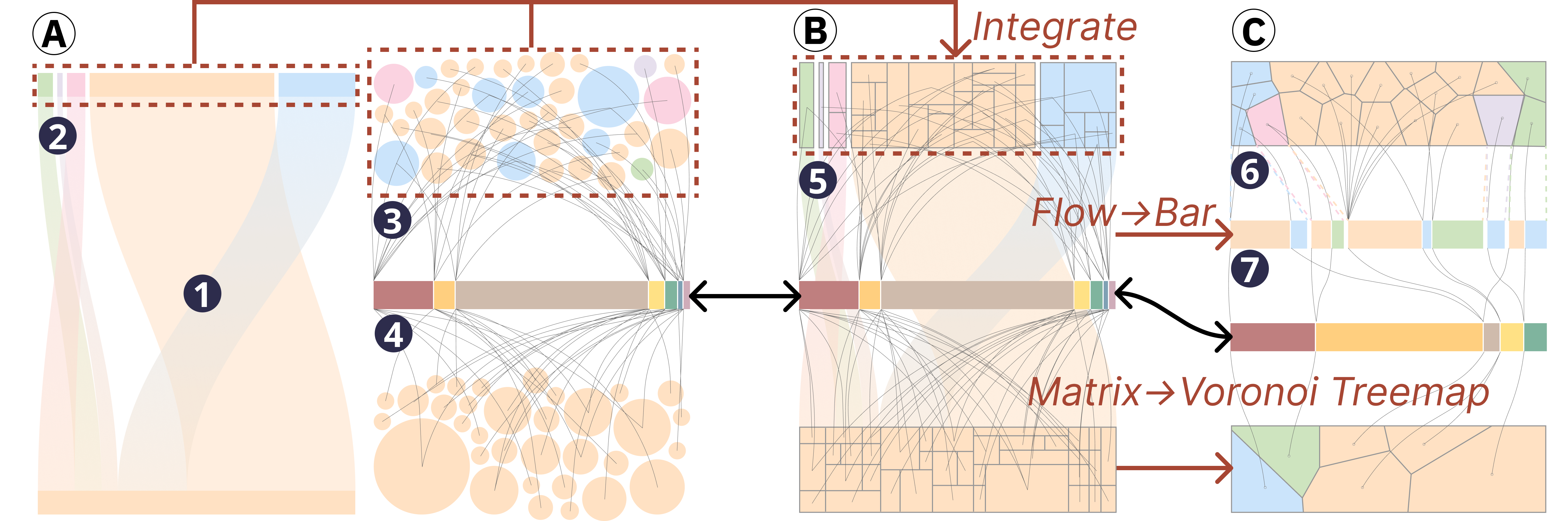}
\vspace{-6mm}
\caption{Design alternatives for \textit{Interpretation View}. \protect\circled{A} is an alternative based on Sankey and Packchart. \protect\circled{B} is an alternative based on Sankey and node matrix. \protect\circled{C} is the final design based on Voronoi Treemap.}
\label{fig:g-alter}
\vspace{-6mm}
\end{figure}

 The \textit{Entity Flow Bar} (\cref{fig:g-alter}-\darkcircled{7}) and the \textit{Path Bar} (\cref{fig:g-alter}-\darkcircled{4}) are both presented as horizontally arranged stacked bars. The \textit{Entity Flow Bar} is positioned beneath the \textit{Entity Treemap} (\cref{fig:g-alter}-\darkcircled{6}) and corresponds to the categories of entities depicted in the \textit{Entity Treemap} above. It illustrates the proportion of the flow from a specific category of entity to the succeeding layer's entity category. The \textit{Entity Flow Bars} for distinct entity categories operate independently, yet entities of the same category are color-coded identically to streamline user comprehension. Flow lines traverse through the designated areas defined by the \textit{Entity Flow Bar}, effectively distinguishing the connections. On the other hand, the \textit{Path Bar} is a comprehensive stacked bar including every type of relation observed, along with its relative proportion. By analyzing these proportions, users can discern the frequency of each relation type. Moreover, connections stemming from the \textit{Entity Flow Bar} are linked to the \textit{Path Bars} representing different edges based on their corresponding relation type, which illustrates the connection between the source entities and the relations.


\par \textbf{Design Alternatives.} In the iterative design process, we explored two alternatives. The first alternative (\cref{fig:g-alter}-\circled{A}) includes the \textit{Sankey Flow} (\cref{fig:g-alter}-\darkcircled{1}) illustrating entity flow, and the \textit{Sankey Bar} (\cref{fig:g-alter}-\darkcircled{2}) showing entity proportion within the interpretation path. In the Packchart diagram, \textit{Entity Nodes} (\cref{fig:g-alter}-\darkcircled{3}) display entity type and proportion, while the \textit{Path bar} (\cref{fig:g-alter}-\darkcircled{4}) shows path type and proportion with connecting lines. This design maintains consistent colors for entity type and accurately represents entity percentage and weight, but confuses hierarchical correspondence between the Sankey and Packchart diagrams, requiring users to toggle between the two. Moreover, the use of nodes in the Packchart is spatially inefficient. In the second alternative (\cref{fig:g-alter}-\circled{B}), we integrate both diagrams by merging \textit{Entity Nodes} and \textit{Sankey Bar} into the \textit{Entity Matrix} (\cref{fig:g-alter}-\darkcircled{5}), where each matrix's size indicates the entity's weight. Retaining \textit{Sankey Flow} and \textit{Path Bar}, this design results in considerable overlap in entities' vertical coordinates, leading to disorganized connections and compromised visibility. Furthermore, the presentation of the hierarchy remains confusing. In the final design (\cref{fig:g-alter}-\circled{C}), \textit{Entity Treemap} (\cref{fig:g-alter}-\darkcircled{6}) eliminates coordinate overlap with irregular shapes, and \textit{Entity Flow Bar} (\cref{fig:g-alter}-\darkcircled{7}) represents the proportion of different entity types, replacing \textit{Sankey Flow}. This design displays hierarchical paths without overlap, ensuring clear and uninterrupted layer connections within an integrated view.

\subsubsection{Modifier View}
\par
\begin{figure*}[h]
\centering
\vspace{-8mm}
\includegraphics[width=\linewidth]{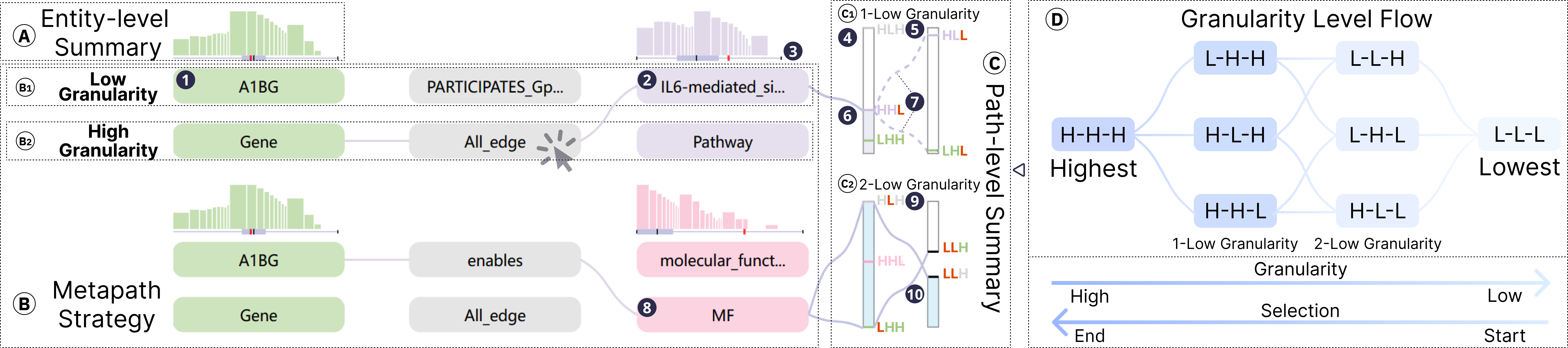}
\vspace{-6mm}
\caption{Metapath Modifier design for Modifier View. \protect\circled{A} Design for the Entity-level Summary. \protect\circled{B} Design for the Metapath Strategy. \protect\circled{C} Design for the Path-level Summary. \protect\circled{D} Granularity Level Flow for the Path-level Summary, H represents High Granularity and L represents Low Granularity.}
\label{fig:d-detail}
\vspace{-6mm}
\end{figure*}

 The \textit{Modifier View} (\cref{fig:pipeline}-\specialcircledg{C}{4}) incorporates both the \textit{Knowledge Graph} (\cref{fig:teaser}-\caserectangle{8}) and \textit{Metapath Modifier} (\cref{fig:d-detail}) functionalities to aid users in exploring the KG (\textbf{DR1}), pinpointing pertinent training data (\textbf{DR2}), and devising human-engaged strategic refinement of metapaths to filter out irrelevant or noisy data within the dataset(\textbf{DR3}).

\par \textbf{Visual Design of the KG}\textbf{.}  Upon selecting a \textit{Primary Gene} in the \textit{Panel}, the KG displays entities originating from this gene. Inspired by the design of \textit{EgoComp}~\cite{liu2016EgoComp}, we used an ego-network layout to show the KG structure. Entities of the same layer are placed on circular arcs with equal radius, preferentially grouping similar entities. \jhr{The colors of entities in the KG correspond to those in the \textit{Interpretation View}, ensuring design consistency.} To manage the exponential increase in entities, we display up to 2-hop connections to balance path length and entity count. When users hover over different entities, the paths leading to that entity and all related entities along the path are highlighted, with their names displayed. This interactive feature allows users to leverage their expertise to evaluate if entities across various paths are biologically related to SL or have plausible links to the \textit{Primary Gene}. \jhr{Automating selection based on interactions with the \textit{Interpretation View} was also considered, but expert feedback showed it could lead users to select many irrelevant paths or paths not present in the KG. Therefore, manual selection was chosen for simplicity and to minimize errors.}

\par \jhr{\textbf{Concept of the Granularity Level Flow.} To assist users in formulating detailed metapath strategies and deriving insights, we introduced the \textit{Granularity Level Flow} (\cref{fig:d-detail}-\circled{D}). This flow illustrates the transition of metapath granularity from \textbf{high (H)} to \textbf{low (L)}, moving left to right, and displays the hierarchy of metapath strategies across granularity levels. In this flow,\coloredboxHHH{H-H-H}represents the metapath with the highest granularity, \textit{1-Low Granularity} represents the metapath with one low granularity entity, and \textit{2-Low Granularity} represents the metapath with two low granularity entity.\coloredboxLLL{L-L-L}denotes specific triplets, with a count of 1. Users begin formulating their metapath strategies by selecting a specific entity from the KG, with the default granularity being\coloredboxLLL{L-L-L}. By converting a low-granularity entity to high-granularity, \jhr{more paths are included in the formulated metapath strategies, resulting in the removal of more paths from the KG due to the expanded selection range.} Therefore, user selections progress from right to left, gradually increasing in granularity. This linear progression helps users determine the appropriate level of granularity, minimizing the risk of accidental or erroneous deletions caused by sudden changes in granularity.}

\par \textbf{Visual Design of the Metapath Modifier.} When a specific entity is selected, the \textit{Metapath Modifier} generates a horizontal metapath display box on the right, linked with the entity. This display box, shaped by usage habits from expert interviews, is intended to show information in a table format. Initially, the display box showcases the selected path (\cref{fig:d-detail}-\specialcircled{B}{1}), with each entity's higher granularity elements (\cref{fig:d-detail}-\specialcircled{B}{2}) displayed beneath it. Higher granularity elements denote wider biological categories, each potentially encompassing several entities of lower granularity. For example, beneath the entity \textit{A1BG}, the metapath element \textit{Gene} is displayed. Users can modify metapath strategies by clicking corresponding entity granularities (\cref{fig:d-detail}-\specialcircled{B}{1} \& \specialcircled{B}{2}). \jhr{A frequency histogram (\cref{fig:d-detail}-\circled{A}) is presented above the source (\cref{fig:d-detail}-\darkcircled{1}) and target entities (\cref{fig:d-detail}-\darkcircled{2}), indicating the distribution of path counts that begin or end with each type of entity. Given the wide range and long-tail distribution of entity frequencies, we applied a log transformation to the x-axis to enhance the comparison of the data, resulting in varying bar widths.} Below the histogram, a horizontal box plot (\cref{fig:d-detail}-\darkcircled{3}) shows the central tendency and dispersion of frequency data. The position of the current entity is marked red, helping users grasp its relative placement within the dataset. To provide detailed information, hovering over a bar in the histogram reveals its interval and corresponding value. To further illustrate the relative proportions of metapath strategies with varying granularity, we have designed two vertical bars (i.e., \textit{Primary Bar} and \textit{Secondary Bar}) (\cref{fig:d-detail}-\darkcircled{4} \& \darkcircled{5}) situated in the \textit{Path-Level Summary} (\cref{fig:d-detail}-\circled{C}). These bars are designed in a manner akin to the \textit{Granularity Level Flow} (\cref{fig:d-detail}-\circled{D}).

\par \textbf{Primary Bar.} The left bar (\cref{fig:d-detail}-\darkcircled{4}), designated as the \textit{Primary Bar}, of which the height corresponds to the number of metapaths with the highest granularity, such as \textit{Gene}---\textit{All\_edge}---\textit{Pathway} (\coloredboxHHH{H-H-H}). Since each entity offers two granularities to choose from---low (\cref{fig:d-detail}-\specialcircled{B}{1}) or high (\cref{fig:d-detail}-\specialcircled{B}{5}), the \textit{Primary Bar} reflects this distinction. Within the \textit{Primary Bar}, three sub-segments (\cref{fig:d-detail}-\darkcircled{6}) depict the relative proportions of metapath strategies with exactly one low granularity entity (i.e.,\coloredboxLHH{L-H-H},\coloredboxLHH{H-L-H}and\coloredboxLHH{H-H-L}) selected within the metapath strategies of highest granularity (i.e.,\coloredboxHHH{H-H-H}), with colors indicating the type of low granularity entity.

\par \textbf{Secondary Bar.} The height of the right bar (\cref{fig:d-detail}-\darkcircled{5}), termed the \textit{Secondary Bar}, corresponds to the number of metapaths involving one low granularity selected entity (i.e.,\coloredboxLHH{L-H-H},\coloredboxLHH{H-L-H}or\coloredboxLHH{H-H-L}). Its sub-segment heights depict the relative proportions of metapath strategies involving two selected low granularity entities within the metapaths of one selected low granularity entity. There are two cases regarding the display of the \textit{Secondary Bar}: 1) If the selected metapath strategy includes exactly one entity of low granularity, the \textit{Secondary Bar} is displayed as a complete bar. Its height represents the number of paths included in the selected metapath strategy, while the sub-segments' height represents the proportion of the two sub-strategies of the selected metapath strategy within it. 2) If the selected metapath strategy includes exactly two entities of low granularity, the \textit{2-Low Granularity} metapath strategies have two parents strategies, such as\coloredboxLLH{L-L-H}among\coloredboxLHH{H-L-H}and\coloredboxLHH{L-H-H}. In this case, the \textit{Secondary Bar} is divided into the top (\cref{fig:d-detail}-\darkcircled{9}) and bottom (\cref{fig:d-detail}-\darkcircled{10}) bars to represent the two proportions, (e.g., \coloredboxLLH{L-L-H}/\coloredboxLHH{H-L-H}and\coloredboxLLH{L-L-H}/\coloredboxLHH{L-H-H}).


\par \textbf{Connecting Rules of \textit{Primary and Secondary Bars} for Granularity Selection.} Based on the number of low granularity entities selected in the strategy, the \textit{Primary Bar} and the \textit{Secondary Bar} utilize two connecting rules to represent corresponding information: \textit{\textbf{1) In cases where exactly one entity of lower granularity is selected}}, such as \textit{Gene}---\textit{All\_edge}---\textit{IL6-mediated\_si...} (\coloredboxLHH{H-H-L}) (\cref{fig:d-detail}-\specialcircled{C}{1}), the target entity (\cref{fig:d-detail}-\darkcircled{2}) links to the corresponding sub-segment (\cref{fig:d-detail}-\darkcircled{6}) on the \textit{Primary Bar} (\cref{fig:d-detail}-\darkcircled{4}). This linkage indicates the strategy's proportion within the highest granularity strategy (\coloredboxHHH{H-H-H}). Two lines (\cref{fig:d-detail}-\darkcircled{7}) extend from the \textit{Primary Bar} to the \textit{Secondary Bar}, delineating the relative proportions of two lower granularity strategies based on the selected entity within the current metapath strategy (\coloredboxLHH{H-H-L}). Since these two lower granularity strategies are not truly selected, they are linked in dash lines. In this example, the \textcolor[RGB]{239, 85, 59}{L} represents the shared low granularity entity, making the \textit{2-Low Granularity} strategies on the \textit{Secondary Bar} subsets of the current strategy. \textit{\textbf{2) In cases where two entities of lower granularity are selected}}, such as \textit{A1BG}---\textit{enables}---\textit{MF} (\coloredboxLLH{L-L-H}) depicted in \cref{fig:d-detail}-\specialcircled{C}{2}, the strategies at this granularity are subsets of two higher-level strategies (\coloredboxLHH{H-L-H}and\coloredboxLHH{L-H-H}), each involving \textit{1-Low Granularity} entity. Consequently, the proportions of these two parent strategies within the highest granularity are simultaneously displayed in the \textit{Primary Bar} and connected to the target entity (\cref{fig:d-detail}-\darkcircled{8}). In this scenario, the \textit{Secondary Bar} (\cref{fig:d-detail}-\darkcircled{9} \& \darkcircled{10}) shows the proportion of the current metapath selection strategy (\coloredboxLLH{L-L-H}) within two higher granularity strategies (\coloredboxLHH{H-L-H}\&\coloredboxLHH{L-H-H}).

\par \textbf{Design Takeaway.} Through observation and interaction with the metapath display box, users can thoroughly assess the proportions and metapaths with specific granularity and derive insights from them. For instance, \textit{A1BG}---\textit{PARTICIPATES}\textit{\_GpPW}---\textit{IL6-MEDIATED\_...} (\cref{fig:d-detail}-\specialcircled{B}{1}), where its \coloredboxLHH{H-H-L} (\cref{fig:d-detail}-\specialcircled{C}{1}) subset occupies a smaller proportion within\coloredboxHHH{H-H-H}, and the \coloredboxLLH{H-L-L} subset occupies a larger proportion within\coloredboxLHH{H-H-L}, user can discern that despite the fewer\coloredboxLHH{H-H-L}paths, the majority of paths between \textit{Gene} (H) and \textit{IL6-MEDIATED\_...} (L) are connected by relation \textit{PARTICIPATES\_GpPW} (L). However, some paths are connected by other relations, like \textit{involved\_in} (L). Gaining this insight, if user determines that \textit{Gene} should not be connected to \textit{IL6-MEDIATED\_...} through \textit{PARTICIPATES\_GpPW}, he can opt for the \coloredboxLLH{H-L-L}metapath instead of\coloredboxLHH{H-H-L}, thereby ensuring the exclusion of any undesired paths from deletion.

\vspace{-1mm}

\section{Evaluation}
\par To evaluate the \textit{SLInterpreter}'s effectiveness, we conducted a case study with \textbf{E1-E3}, who had participated in our formative study. Following the Co-discovery Learning Protocol~\cite{lim1997empirical}, one author briefly guided the system's operation while \textbf{E1} operated the system and \textbf{E2} and \textbf{E3} engaged in discussions. We then conducted expert interviews to gather insights and feedback on their user experience. \jhr{The entire process lasted about 2 hours: 30 minutes for familiarizing participants with the system, followed by two 50-minute rounds of iterative exploration.}

\vspace{-1mm}

\subsection{Case Study}
\par
\begin{wrapfigure}{r}{0.3\columnwidth}
 \vspace{-3mm}
 \centering 
 \includegraphics[width=0.3\columnwidth]{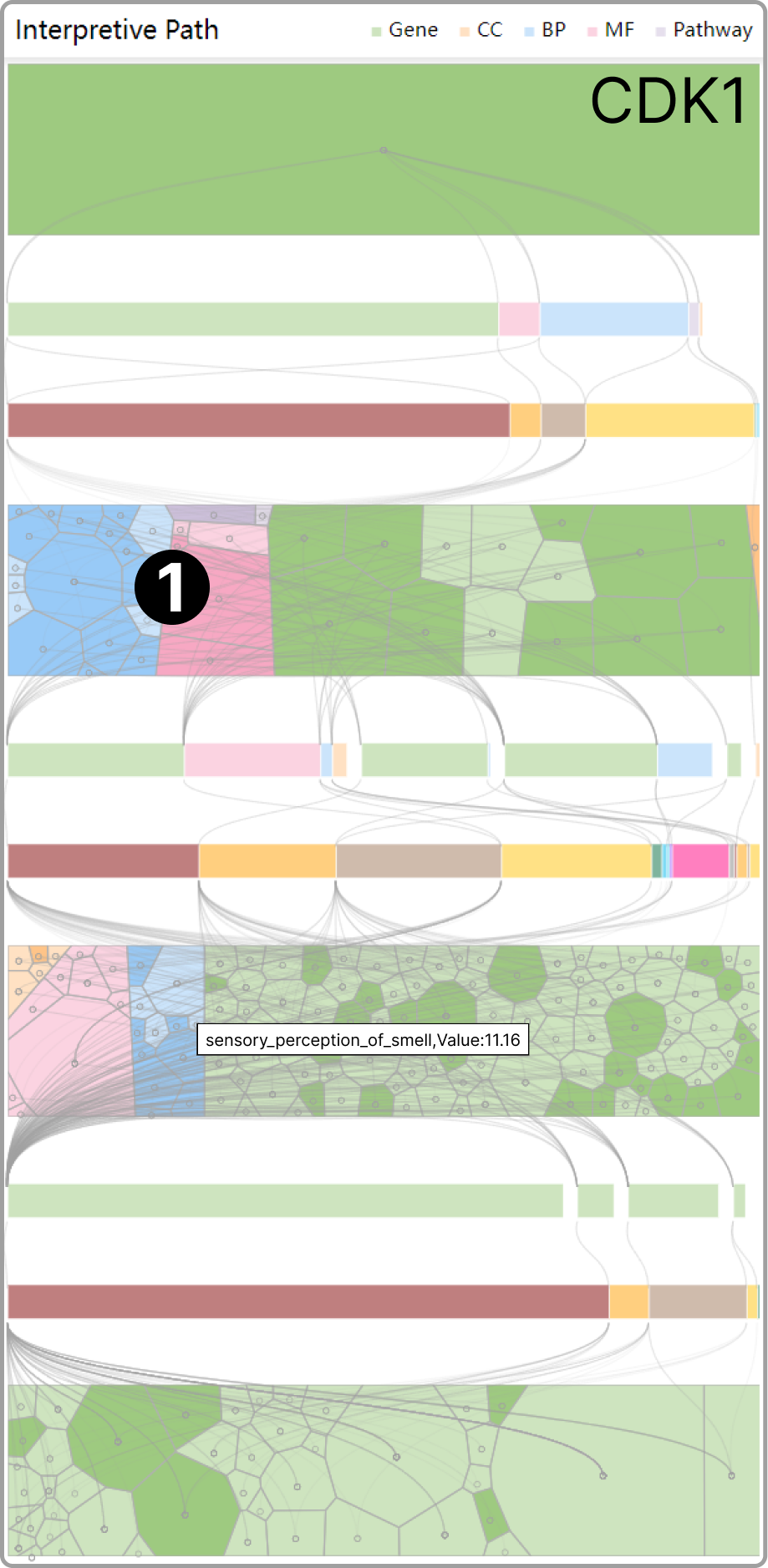}
 \vspace{-6mm}
 \caption{\protect\darkcircled{1} \textbf{E1} clicked on the non-gene entities during investigating CDK1.}
 \label{fig:case2_1}
  \vspace{-6mm}
\end{wrapfigure}

 \textbf{Investigating CDK1 and Identifying Problem.} \textbf{E1}, a thyroid cancer specialist, initiated by selecting thyroid cancer in the \textit{Disease search box} (\cref{fig:teaser}-\caserectangle{1}). Within the \textit{Embedding View}, he identified a region (\cref{fig:teaser}-\caserectangle{2}) where accurately predicted \textit{Partner Genes} are closely clustered. Employing semantic zoom, \textbf{E1} noticed a cluster of KRAS predictions in the upper right area. Given the extensive study of KRAS in SL and its role in the RAS/MAPK signaling pathway regulating cells, the experts held a solid understanding of KRAS. Observing that correctly predicted \textit{Partner Genes} of CDK1 also cluster in the upper right, similar to KRAS, \textbf{E1} selected CDK1 in the \textit{Primary Gene search box} (\cref{fig:teaser}-\caserectangle{3}). However, he found the interpretive paths for CDK1 predominantly linked by gene, making these paths less convincing. Therefore, \textbf{E1} shifted focus to different path types. By clicking on the non-gene entities (\cref{fig:case2_1}-\darkcircled{1}), \textbf{E1} discovered that these paths primarily lead to MYC and RPL13. Among the non-gene entities, \textit{sensory\_perception\_of\_smell} has the highest weight in BP, which \textbf{E2} deemed irrelevant to SL.

\par \textbf{Analysing the Interpretive Path of \textit{CDK1$\rightarrow$MYC}.} Delving deeper into the embedding of \textit{CDK1}, \textbf{E1} noted three correctly predicted genes: \textit{KRAS}, ranked $13^{th}$, \textit{MYC} at $27^{th}$, and \textit{RPL13} at $33^{rd}$. \jhr{After excluding higher-ranked results that were predicted solely based on SL relationships and had lower biological significance, \textbf{E1} found the remaining ranks sufficiently high to merit further exploration.} Notably, \textit{CDK1} and \textit{KRAS} are close in the embedding, as are \textit{MYC} and \textit{RPL13}. This proximity suggests a similarity in their features. Given their relevance to thyroid cancer, the clustering of \textit{CDK1} and \textit{KRAS} is intriguing to \textbf{E1}. He emphasized that targeted therapy for \textit{RAS} gene mutations is a key focus in cancer treatment, with extensive research ongoing. Consequently, \textbf{E1} was interested in how the model explains the predictions for the other two genes. Drawing from his previous observations, \textbf{E1} proceeded by selecting the higher-ranked \textit{MYC} in the \textit{Partner Gene Table} (\cref{fig:teaser}-\caserectangle{4}), and the corresponding tag is generated (\cref{fig:teaser}-\caserectangle{5}). \textit{MYC} is a critical oncogene encoding the transcription factor \textit{MYC} protein, pivotal in cell growth, division, and metabolism processes. Analyzing the \textit{CDK1$\rightarrow$MYC} interpretive path, \textbf{E1} observed that the \textit{Entity Treemap}'s middle layers primarily consist of \textit{Genes}, \textit{BP}, and \textit{CC}. Contrary to common patterns observed in \textit{CDK1}'s pairing with \textit{Partner Genes}, the \textit{Entity Flow Bar} (\cref{fig:teaser}-\caserectangle{6}) indicates that gene-gene connections are not predominant here. Additionally, the \textit{Path Bar} (\cref{fig:teaser}-\caserectangle{7}) displays a smaller ratio of \textit{SL\_GsG}, which indicates validated SL partner relationship, compared to those belonging to BP and CC. The linkage between \textit{CDK1} and \textit{MYC} mainly involves \textit{Gene$\rightarrow$BP} paths, such as \textit{CDK1---involved\_in---DNA\_replication---has\_part---DNA\_biosynthetic\_process---involved\_in\_inv---MYC}. \textbf{E1} found this path consistent with his domain knowledge, particularly as \textit{MYC} overexpression, a transcription factor driving cell growth, may enhance the reliance on \textit{CDK1} for DNA biosynthesis.





\par \textbf{Refining the KG.} From the initial exploration, \textbf{E2} identified an implausible path \textit{CDK1---SL\_GsG---FARSA---involved\_in---sensory\_perception\_of\_smell---involved\_in\_inv---MYC} and similar paths inconsistent with his expertise. To investigate potential data-level issues causing these irrational predictions, he employed the \textit{Modifier View}. \jhr{\textbf{E1} first narrowed the investigation scope to \textit{CDK1$\rightarrow$Gene$\rightarrow$BP} in the \textit{Knowledge Graph}. Since \textit{CDK1} belongs to the gene category, he began exploring from the dashed green areas (\cref{fig:teaser}-\caserectangle{8})}. He noticed that the path of \textit{Gene$\rightarrow$sensory\_perception\_of\_smell} indeed appears in the \textit{Knowledge Graph}, despite its irrelevance to SL. Selecting this path in the \textit{Knowledge Graph} (\cref{fig:teaser}-\caserectangle{9}), \textbf{E1} then refined metapath strategies in the \textit{Metapath Modifier}. He first noticed the histogram of \textit{sensory\_perception\_of\_smell} (\cref{fig:teaser}-\caserectangle{10}), which indicates that among all entities in its high granularity category of \textit{Biological Processes (BP)}, this entity frequently appears as the endpoint of such metapaths (\textit{Gene}---\textit{All\_edge}---\textit{BP}). Based on his expertise, \textbf{E1} determined that any path involving \textit{sensory\_perception\_of\_smell} is irrelevant to SL. Therefore, \textbf{E1} selected the \coloredboxLLH{L-H-L} strategy by clicking on \textit{All edge} (\cref{fig:teaser}-\caserectangle{11}).

\par \jhr{\textbf{E3} noted the relatively low height of the primary bar, indicating that the proportional relationship of \textit{FARSA---All edge---sensory\_perception\_of\_smell} (\coloredboxLHH{L-H-L}) within its parent set, \textit{Gene---All edge---sensory\_perception\_of\_smell} (\coloredboxLHH{H-H-L}) (\cref{fig:teaser}-\caserectangle{12}), was not statistically significant.} This indicates that many other \textit{Genes} are also connected to \textit{sensory\_perception\_of\_smell}, which are all irrelevant to SL. Consequently, \textbf{E1} selected the \coloredboxLHH{H-H-L} metapath strategy by clicking on \textit{Gene} (\cref{fig:teaser}-\caserectangle{13}). Upon doing so, \textbf{E1} observed the proportional relationships of the \coloredboxLLH{L-H-L} and \coloredboxLLH{H-L-L} sub-strategies to \coloredboxLHH{H-H-L} (\cref{fig:teaser}-\caserectangle{14}). Specifically, there is only one occurrence of \coloredboxLLH{L-H-L}, while \coloredboxLLH{H-L-L} dominates the entire bar. This suggests that almost all \textit{Genes} are associated with \textit{sensory\_perception\_of\_smell} through the \textit{involved\_in} relationship. Simultaneously, FARSA represents only a fraction of \textit{Gene} connected to \textit{sensory\_perception\_of\_smell}, even when considering all possible edge types. These findings aligned with the experts' expectations. Consequently, \textbf{E1} added this strategy to the \textit{Operation List} and entered ``\textit{sensory\_perception\_of\_smell is irrelevant to synthetic lethality}'' in the text box (\cref{fig:teaser}-\caserectangle{15}). Subsequently, \textbf{E1} clicked \raisebox{-0.4ex}{\includegraphics[height=2ex]{figs/Group-496.png}} (\cref{fig:teaser}-\caserectangle{16}) to retrain the model. Upon receiving the retrained model and its performance metrics from the backend, \textbf{E1} observed from the \textit{Model Log} that the new model's performance was slightly improved compared to the original model. The accuracy increased by $0.88\%$, an $1.66\%$ improvement over the original rate, while the deleted path data accounted for only $0.13\%$ of the total data.

\begin{wrapfigure}{r}{0.3\columnwidth}
 \vspace{-3mm}
 \centering 
 \includegraphics[width=0.3\columnwidth]{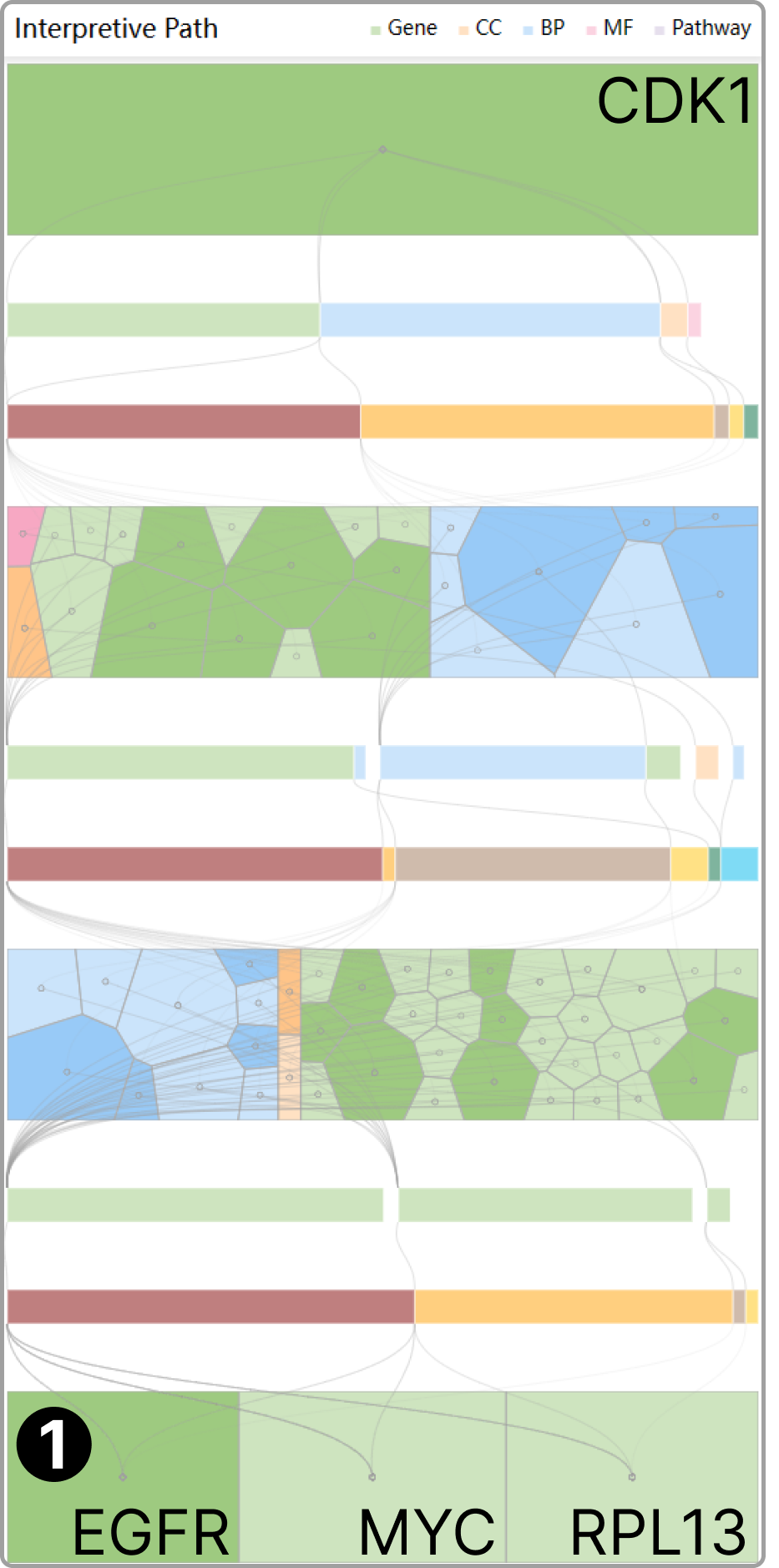}
 \vspace{-6mm}
 \caption{\protect\darkcircled{1} \textbf{E1} highlighted EGFR’s interpretive paths to reveal patterns similar to MYC and RPL13.}
 \label{fig:case2_2}
  \vspace{-6mm}
\end{wrapfigure}

\par \textbf{Exploring New Potential SL Partner of CDK1.} \textbf{E1} proceeded by selecting the retrained model(\cref{fig:teaser}-\caserectangle{17}) to initiate an iterative exploration. This time, he directly targeted CDK1 in the \textit{Primary Gene search box}. Upon examining the \textit{Interpretion View} of CDK1, \textbf{E2} noticed that the newly generated paths no longer include \textit{sensory\_perception\_of\_smell}. Furthermore, MYC and RPL13 remained accurately identified, maintaining their rankings at $23^{rd}$ and $29^{th}$. In the \textit{Embedding View}, \textbf{E1} noted that RPL13 and EGFR are close to MYC. Additionally, through the \textit{Rank Indicator} (\cref{fig:teaser}-\caserectangle{18}), \textbf{E3} noticed that these three genes share similar rankings. Notably, EGFR, ranked $31^{st}$, indicates a potential but unidentified SL relationship with CDK1. To investigate potential patterns among them, \textbf{E1} conducted a lasso operation (\cref{fig:teaser}-\caserectangle{19}). Among these genes' interpretive paths, \textbf{E3} observed a high proportion for the pattern \textit{Gene$\rightarrow$BP$\rightarrow$BP$\rightarrow$Gene}. By highlighting EGFR's interpretive paths (\cref{fig:case2_2}-\darkcircled{1}), \textbf{E3} observed striking similarities to the patterns of the other two correctly predicted genes. At this point, \textbf{E3} remarked, ``\textit{If the model could give us clear paths like this for every prediction, and we could compare them with other known patterns, then I'd feel more confident in trusting those predictions and putting resources into checking them.}

\vspace{-1mm}

\subsection{Expert Interview}

\par \textbf{System Design and Feedback.} Biologists confirmed the effectiveness of \textit{SLInterpreter} in a human-AI collaboration that enhances model interpretability and explores SL mechanisms. \textbf{E1} remarked, ``\textit{The design of the \textit{Session View} fits well with my habits of using biological databases; we can start exploring from a specific disease of interest.}'' \textbf{E2} praised the \textit{Embedding View} for quickly indicating potential common patterns and appreciated the \textit{Rank Indicator}, which helps determine the ranks of predicted genes. \textbf{E3} was pleased with the system's comprehensive and intuitive features that aid in removing irrelevant or erroneous data, especially given the high number of data proven to be erroneous or biased. He suggested the system could further improve by automatically collecting and displaying suspicious data. \jhr{Additionally, \textbf{E1} and \textbf{E3} noted that traditional AI-assisted methods required sequential reviewing of a large number of materials based on confidence ranking alone, typically taking \textbf{two to three days} to filter $50$ predicted results. This lengthy duration was due to the extensive time spent retrieving potential associations between predicted SL pairs, as the information was scattered and lacked targeted retrieval, often resulting in unexpected delays. The discontinuity of this review process also reduced their confidence in the results. In contrast, using \textit{SLInterpreter} made their exploration more targeted, enabling them to review $50$ predicted results within \textbf{five to eight hours}, considering the varying time required for retraining the model. \textit{SLInterpreter} eliminated the need for further retrieval of obviously erroneous prediction paths, and the continuity and intervenability of the process increased their confidence in the results.}

\noindent\textbf{Learning Curve.} The experts noted that the system's design matches their usage habits and the interaction is intuitive. However, there is a certain learning curve associated with the \textit{Metapath Modifier} in the \textit{Modifier View} and the whole iterative cycle. The entire system requires some time to understand and master (about 20 minutes), and learning through specific cases can make the learning curve more gradual. Nevertheless, the experts still believe that \textit{SLInterpreter} can help them gain more insights from their explorations and have greater confidence in the potential SL pairs they discover.

\vspace{-1mm}

\section{Discussion and Limitation}
\noindent\textbf{Lessons Learned.} Our collaboration with biologists has provided valuable insights. First, SL-related biological data is often extensive and high-dimensional, making rule-based filtering insufficient and necessitating biologists' expertise. However, exhaustive evaluations by biologists are unsustainable, highlighting the importance of leveraging AI or visualization tools to focus their assessments. Second, we recognized the need to refine metapath granularity levels. While biological knowledge categorization is intricate, adding granularity levels for certain entity types can rapidly expand metapath strategies. Thus, balancing granularity levels with user burden is crucial for effeteness.

\noindent\textbf{Generalization and Scalability.} \jhr{The \textit{SLInterpreter} enhances interpretability and facilitates cross-granularity exploration for AI models through interpretive paths. Validated by biologists, it is applicable to various network predictive tasks and models across multiple domains, including drug prediction, social network forecasting, and recommendation systems. To ensure scalability, features like semantic zooming and highlighting have been implemented to prevent visual clutter. Additionally, for larger entity sets in knowledge graphs, stacking or grouping nodes can further enhance spatial information capacity.}

\noindent\textbf{Limitations.} This study presents several limitations. First, the dataset has limited biological relationships. Despite SL spans various domains like biomolecular studies and genomics, resulting explanations remain somewhat broad and constrained. Second, the model primarily predicts new SL pairs via existing pairs, which may have limited value due to relatively few connected paths via biological information. Third, experts note that SL gene pairs are not exclusively limited to pairs~\cite{ryan2023complex}. \jhr{Although rare, confirmed instances of SL relationships involving multiple genes exist, expanding prediction from two genes to more, complicating their connections.} This highlights the need for comprehensive systems and visualizations for complex SL networks. \jhr{Additionally, due to extended cycles of wet lab experiments, adequate quantitative experiments were not feasible within the available time.}

\vspace{-1mm}

\section{Conclusion and Future Work}
\par This study introduces an iterative Human-AI collaboration framework using \textit{SLInterpreter}, aimed at \textit{1) Human-Engaged Knowledge Graph Refinement based on Metapath Strategies} and \textit{2) Cross-Granularity SL Interpretation Enhancement and Mechanism Analysis} for domain experts. A case study and expert interviews demonstrate \textit{SLInterpreter}'s ability to efficiently discover new SL pairs, provide substantial interpretability, and offer effective methods for commonality exploration. \jhr{Future plans include extending the analysis from single paths to complex networks involving multiple genes, further integrating additional data sources, and expanding the system to other GNN prediction tasks.}

\acknowledgments{%
We would like to express our gratitude to our domain experts and the anonymous reviewers for their insightful comments. This work is funded by grants from the National Natural Science Foundation of China (No. 62372298), the Shanghai Frontiers Science Center of Human-centered Artificial Intelligence (ShangHAI), and the Key Laboratory of Intelligent Perception and Human-Machine Collaboration (ShanghaiTech University), Ministry of Education.
}

\bibliographystyle{abbrv-doi-hyperref}

\balance
\bibliography{template}

\begin{thebibliography}{10}

\bibitem{ashktorab2020human}
Z.~Ashktorab, Q.~V. Liao, C.~Dugan, J.~Johnson, Q.~Pan, W.~Zhang, S.~Kumaravel, and M.~Campbell.
\newblock Human-ai collaboration in a cooperative game setting: Measuring social perception and outcomes.
\newblock {\em Proceedings of the ACM on Human-Computer Interaction}, 4(CSCW2):1--20, 2020. \href{https://doi.org/10.1145/3415167}
{doi: {{%
10\hspace{.1pt}\discretionary{.}{%
}{.}\hspace{.4pt}1145\discretionary{/}{%
}{/}3415167}}}


\bibitem{baldassarre2019explainability}
F.~Baldassarre and H.~Azizpour.
\newblock Explainability techniques for graph convolutional networks.
\newblock {\em arXiv preprint arXiv:1905.13686}, 2019. \href{https://doi.org/10.48550/arXiv.1905.1368}
{doi: {{%
10\hspace{.1pt}\discretionary{.}{%
}{.}\hspace{.4pt}48550\discretionary{/}{%
}{/}arXiv\hspace{.1pt}\discretionary{.}{%
}{.}\hspace{.4pt}1905\hspace{.1pt}\discretionary{.}{%
}{.}\hspace{.4pt}1368}}}


\bibitem{balzer2005Voronoi}
M.~Balzer and O.~Deussen.
\newblock Voronoi treemaps.
\newblock In {\em IEEE Symposium on Information Visualization, 2005. INFOVIS 2005.}, pp. 49--56, 2005. \href{https://doi.org/10.1109/INFVIS.2005.1532128}
{doi: {{%
10\hspace{.1pt}\discretionary{.}{%
}{.}\hspace{.4pt}1109\discretionary{/}{%
}{/}INFVIS\hspace{.1pt}\discretionary{.}{%
}{.}\hspace{.4pt}2005\hspace{.1pt}\discretionary{.}{%
}{.}\hspace{.4pt}1532128}}}


\bibitem{cai2020dual}
R.~Cai, X.~Chen, Y.~Fang, M.~Wu, and Y.~Hao.
\newblock Dual-dropout graph convolutional network for predicting synthetic lethality in human cancers.
\newblock {\em Bioinformatics}, 36(16):4458--4465, 2020. \href{https://doi.org/10.1093/bioinformatics/btaa211}
{doi: {{%
10\hspace{.1pt}\discretionary{.}{%
}{.}\hspace{.4pt}1093\discretionary{/}{%
}{/}bioinformatics\discretionary{/}{%
}{/}btaa211}}}


\bibitem{clarke2017thematic}
V.~Clarke and V.~Braun.
\newblock Thematic analysis.
\newblock {\em The journal of positive psychology}, 12(3):297--298, 2017. \href{https://doi.org/10.1080/17439760.2016.1262613}
{doi: {{%
10\hspace{.1pt}\discretionary{.}{%
}{.}\hspace{.4pt}1080\discretionary{/}{%
}{/}17439760\hspace{.1pt}\discretionary{.}{%
}{.}\hspace{.4pt}2016\hspace{.1pt}\discretionary{.}{%
}{.}\hspace{.4pt}1262613}}}


\bibitem{conde2009human}
N.~Conde-Pueyo, A.~Munteanu, R.~V. Sol{\'e}, and C.~Rodr{\'\i}guez-Caso.
\newblock Human synthetic lethal inference as potential anti-cancer target gene detection.
\newblock {\em BMC Systems Biology}, 3(1):1--15, 2009. \href{https://doi.org/10.1186/1752-0509-3-116}
{doi: {{%
10\hspace{.1pt}\discretionary{.}{%
}{.}\hspace{.4pt}1186\discretionary{/}{%
}{/}1752\discretionary{%
}{-}{-}0509\discretionary{%
}{-}{-}3\discretionary{%
}{-}{-}116}}}


\bibitem{deng2019sl}
X.~Deng, S.~Das, K.~Valdez, K.~Camphausen, and U.~Shankavaram.
\newblock Sl-biodp: multi-cancer interactive tool for prediction of synthetic lethality and response to cancer treatment.
\newblock {\em Cancers}, 11(11):1682, 2019. \href{https://doi.org/10.3390/cancers11111682}
{doi: {{%
10\hspace{.1pt}\discretionary{.}{%
}{.}\hspace{.4pt}3390\discretionary{/}{%
}{/}cancers11111682}}}


\bibitem{dudley2018review}
J.~J. Dudley and P.~O. Kristensson.
\newblock A review of user interface design for interactive machine learning.
\newblock {\em ACM Transactions on Interactive Intelligent Systems (TiiS)}, 8(2):1--37, 2018. \href{https://doi.org/10.1145/3185517}
{doi: {{%
10\hspace{.1pt}\discretionary{.}{%
}{.}\hspace{.4pt}1145\discretionary{/}{%
}{/}3185517}}}


\bibitem{feng2019platform}
X.~Feng, N.~Arang, D.~C. Rigiracciolo, J.~S. Lee, H.~Yeerna, Z.~Wang, S.~Lubrano, A.~Kishore, J.~A. Pachter, G.~M. K{\"o}nig, et~al.
\newblock A platform of synthetic lethal gene interaction networks reveals that the gnaq uveal melanoma oncogene controls the hippo pathway through fak.
\newblock {\em Cancer cell}, 35(3):457--472, 2019. \href{https://doi.org/10.1016/j.ccell.2019.01.009}
{doi: {{%
10\hspace{.1pt}\discretionary{.}{%
}{.}\hspace{.4pt}1016\discretionary{/}{%
}{/}j\hspace{.1pt}\discretionary{.}{%
}{.}\hspace{.4pt}ccell\hspace{.1pt}\discretionary{.}{%
}{.}\hspace{.4pt}2019\hspace{.1pt}\discretionary{.}{%
}{.}\hspace{.4pt}01\hspace{.1pt}\discretionary{.}{%
}{.}\hspace{.4pt}009}}}


\bibitem{geer2010ncbi}
L.~Y. Geer, A.~Marchler-Bauer, R.~C. Geer, L.~Han, J.~He, S.~He, C.~Liu, W.~Shi, and S.~H. Bryant.
\newblock The ncbi biosystems database.
\newblock {\em Nucleic acids research}, 38(suppl\_1):D492--D496, 2010. \href{https://doi.org/10.1093/nar/gkp858}
{doi: {{%
10\hspace{.1pt}\discretionary{.}{%
}{.}\hspace{.4pt}1093\discretionary{/}{%
}{/}nar\discretionary{/}{%
}{/}gkp858}}}


\bibitem{ghandi2014kmer}
M.~Ghandi, D.~Lee, M.~Mohammad-Noori, and M.~A. Beer.
\newblock Enhanced regulatory sequence prediction using gapped k-mer features.
\newblock {\em PLOS Computational Biology}, 10(7):1--15, 07 2014. \href{https://doi.org/10.1371/journal.pcbi.1003711}
{doi: {{%
10\hspace{.1pt}\discretionary{.}{%
}{.}\hspace{.4pt}1371\discretionary{/}{%
}{/}journal\hspace{.1pt}\discretionary{.}{%
}{.}\hspace{.4pt}pcbi\hspace{.1pt}\discretionary{.}{%
}{.}\hspace{.4pt}1003711}}}


\bibitem{gourley2019moving}
C.~Gourley, J.~Balma{\~n}a, J.~A. Ledermann, V.~Serra, R.~Dent, S.~Loibl, E.~Pujade-Lauraine, and S.~J. Boulton.
\newblock Moving from poly (adp-ribose) polymerase inhibition to targeting dna repair and dna damage response in cancer therapy.
\newblock {\em Journal of Clinical Oncology}, 37(25):2257--2269, 2019. \href{https://doi.org/10.1200/JCO.18.02050}
{doi: {{%
10\hspace{.1pt}\discretionary{.}{%
}{.}\hspace{.4pt}1200\discretionary{/}{%
}{/}JCO\hspace{.1pt}\discretionary{.}{%
}{.}\hspace{.4pt}18\hspace{.1pt}\discretionary{.}{%
}{.}\hspace{.4pt}02050}}}


\bibitem{huang2023concept}
J.~Huang, A.~Mishra, B.~C. Kwon, and C.~Bryan.
\newblock Conceptexplainer: Interactive explanation for deep neural networks from a concept perspective.
\newblock {\em IEEE Transactions on Visualization and Computer Graphics}, 29(1):831--841, 2023. \href{https://doi.org/10.1109/TVCG.2022.3209384}
{doi: {{%
10\hspace{.1pt}\discretionary{.}{%
}{.}\hspace{.4pt}1109\discretionary{/}{%
}{/}TVCG\hspace{.1pt}\discretionary{.}{%
}{.}\hspace{.4pt}2022\hspace{.1pt}\discretionary{.}{%
}{.}\hspace{.4pt}3209384}}}


\bibitem{ij2018statistics}
H.~Ij.
\newblock Statistics versus machine learning.
\newblock {\em Nat Methods}, 15(4):233, 2018. \href{https://doi.org/10.1038/nmeth.4642}
{doi: {{%
10\hspace{.1pt}\discretionary{.}{%
}{.}\hspace{.4pt}1038\discretionary{/}{%
}{/}nmeth\hspace{.1pt}\discretionary{.}{%
}{.}\hspace{.4pt}4642}}}


\bibitem{jacunski2015connectivity}
A.~Jacunski, S.~J. Dixon, and N.~P. Tatonetti.
\newblock Connectivity homology enables inter-species network models of synthetic lethality.
\newblock {\em PLoS computational biology}, 11(10):e1004506, 2015. \href{https://doi.org/10.1371/journal.pcbi.1004506}
{doi: {{%
10\hspace{.1pt}\discretionary{.}{%
}{.}\hspace{.4pt}1371\discretionary{/}{%
}{/}journal\hspace{.1pt}\discretionary{.}{%
}{.}\hspace{.4pt}pcbi\hspace{.1pt}\discretionary{.}{%
}{.}\hspace{.4pt}1004506}}}


\bibitem{jakubik2022empirical}
J.~Jakubik, J.~Sch{\"o}ffer, V.~Hoge, M.~V{\"o}ssing, and N.~K{\"u}hl.
\newblock An empirical evaluation of predicted outcomes as explanations in human-ai decision-making.
\newblock In {\em Joint European Conference on Machine Learning and Knowledge Discovery in Databases}, pp. 353--368. Springer, 2022. \href{https://doi.org/10.48550/arXiv.2208.04181}
{doi: {{%
10\hspace{.1pt}\discretionary{.}{%
}{.}\hspace{.4pt}48550\discretionary{/}{%
}{/}arXiv\hspace{.1pt}\discretionary{.}{%
}{.}\hspace{.4pt}2208\hspace{.1pt}\discretionary{.}{%
}{.}\hspace{.4pt}04181}}}


\bibitem{SLSTPMC}
H.~Jariyal, F.~Weinberg, A.~Achreja, D.~Nagarath, and A.~Srivastava.
\newblock Synthetic lethality: a step forward for personalized medicine in cancer.
\newblock {\em Drug Discovery Today}, 25(2):305--320, 2020. \href{https://doi.org/10.1016/j.drudis.2019.11.014}
{doi: {{%
10\hspace{.1pt}\discretionary{.}{%
}{.}\hspace{.4pt}1016\discretionary{/}{%
}{/}j\hspace{.1pt}\discretionary{.}{%
}{.}\hspace{.4pt}drudis\hspace{.1pt}\discretionary{.}{%
}{.}\hspace{.4pt}2019\hspace{.1pt}\discretionary{.}{%
}{.}\hspace{.4pt}11\hspace{.1pt}\discretionary{.}{%
}{.}\hspace{.4pt}014}}}


\bibitem{jerby2014predicting}
L.~Jerby-Arnon, N.~Pfetzer, Y.~Y. Waldman, L.~McGarry, D.~James, E.~Shanks, B.~Seashore-Ludlow, A.~Weinstock, T.~Geiger, P.~A. Clemons, et~al.
\newblock Predicting cancer-specific vulnerability via data-driven detection of synthetic lethality.
\newblock {\em Cell}, 158(5):1199--1209, 2014. \href{https://doi.org/10.1016/j.cell.2014.07.027}
{doi: {{%
10\hspace{.1pt}\discretionary{.}{%
}{.}\hspace{.4pt}1016\discretionary{/}{%
}{/}j\hspace{.1pt}\discretionary{.}{%
}{.}\hspace{.4pt}cell\hspace{.1pt}\discretionary{.}{%
}{.}\hspace{.4pt}2014\hspace{.1pt}\discretionary{.}{%
}{.}\hspace{.4pt}07\hspace{.1pt}\discretionary{.}{%
}{.}\hspace{.4pt}027}}}


\bibitem{1339264}
D.~Jiang, C.~Tang, and A.~Zhang.
\newblock Cluster analysis for gene expression data: a survey.
\newblock {\em IEEE Transactions on Knowledge and Data Engineering}, 16(11):1370--1386, 2004. \href{https://doi.org/10.1109/TKDE.2004.68}
{doi: {{%
10\hspace{.1pt}\discretionary{.}{%
}{.}\hspace{.4pt}1109\discretionary{/}{%
}{/}TKDE\hspace{.1pt}\discretionary{.}{%
}{.}\hspace{.4pt}2004\hspace{.1pt}\discretionary{.}{%
}{.}\hspace{.4pt}68}}}


\bibitem{jin2022gnnlens}
Z.~Jin, Y.~Wang, Q.~Wang, Y.~Ming, T.~Ma, and H.~Qu.
\newblock Gnnlens: A visual analytics approach for prediction error diagnosis of graph neural networks.
\newblock {\em IEEE Transactions on Visualization and Computer Graphics}, 2022. \href{https://doi.org/10.1109/TVCG.2022.3148107}
{doi: {{%
10\hspace{.1pt}\discretionary{.}{%
}{.}\hspace{.4pt}1109\discretionary{/}{%
}{/}TVCG\hspace{.1pt}\discretionary{.}{%
}{.}\hspace{.4pt}2022\hspace{.1pt}\discretionary{.}{%
}{.}\hspace{.4pt}3148107}}}


\bibitem{TCSLCAT}
W.~G. Kaelin~Jr.
\newblock The concept of synthetic lethality in the context of anticancer therapy.
\newblock {\em Nature reviews cancer}, 5(9):689--698, 2005. \href{https://doi.org/10.1038/nrc1691}
{doi: {{%
10\hspace{.1pt}\discretionary{.}{%
}{.}\hspace{.4pt}1038\discretionary{/}{%
}{/}nrc1691}}}


\bibitem{kranthi2013identification}
T.~Kranthi, S.~Rao, and P.~Manimaran.
\newblock Identification of synthetic lethal pairs in biological systems through network information centrality.
\newblock {\em Molecular bioSystems}, 9(8):2163--2167, 2013. \href{https://doi.org/10.1039/c3mb25589a}
{doi: {{%
10\hspace{.1pt}\discretionary{.}{%
}{.}\hspace{.4pt}1039\discretionary{/}{%
}{/}c3mb25589a}}}


\bibitem{ku2020integration}
A.~A. Ku, H.-M. Hu, X.~Zhao, K.~N. Shah, S.~Kongara, D.~Wu, F.~McCormick, A.~Balmain, and S.~Bandyopadhyay.
\newblock Integration of multiple biological contexts reveals principles of synthetic lethality that affect reproducibility.
\newblock {\em Nature communications}, 11(1):2375, 2020. \href{https://doi.org/10.1038/s41467-020-16078-y}
{doi: {{%
10\hspace{.1pt}\discretionary{.}{%
}{.}\hspace{.4pt}1038\discretionary{/}{%
}{/}s41467\discretionary{%
}{-}{-}020\discretionary{%
}{-}{-}16078\discretionary{%
}{-}{-}y}}}


\bibitem{PSLHCMGENN}
M.~Lai, G.~Chen, H.~Yang, J.~Yang, Z.~Jiang, M.~Wu, and J.~Zheng.
\newblock Predicting synthetic lethality in human cancers via multi-graph ensemble neural network.
\newblock In {\em 2021 43rd Annual International Conference of the IEEE Engineering in Medicine \& Biology Society (EMBC)}, pp. 1731--1734, 2021. \href{https://doi.org/10.1109/EMBC46164.2021.9630716}
{doi: {{%
10\hspace{.1pt}\discretionary{.}{%
}{.}\hspace{.4pt}1109\discretionary{/}{%
}{/}EMBC46164\hspace{.1pt}\discretionary{.}{%
}{.}\hspace{.4pt}2021\hspace{.1pt}\discretionary{.}{%
}{.}\hspace{.4pt}9630716}}}


\bibitem{lai2022human}
V.~Lai, S.~Carton, R.~Bhatnagar, Q.~V. Liao, Y.~Zhang, and C.~Tan.
\newblock Human-ai collaboration via conditional delegation: A case study of content moderation.
\newblock In {\em Proceedings of the 2022 CHI Conference on Human Factors in Computing Systems}, pp. 1--18, 2022. \href{https://doi.org/10.1145/3491102.3501999}
{doi: {{%
10\hspace{.1pt}\discretionary{.}{%
}{.}\hspace{.4pt}1145\discretionary{/}{%
}{/}3491102\hspace{.1pt}\discretionary{.}{%
}{.}\hspace{.4pt}3501999}}}


\bibitem{lai2020chicago}
V.~Lai, H.~Liu, and C.~Tan.
\newblock " why is' chicago'deceptive?" towards building model-driven tutorials for humans.
\newblock In {\em Proceedings of the 2020 CHI Conference on Human Factors in Computing Systems}, pp. 1--13, 2020. \href{https://doi.org/10.1145/3313831.3376873}
{doi: {{%
10\hspace{.1pt}\discretionary{.}{%
}{.}\hspace{.4pt}1145\discretionary{/}{%
}{/}3313831\hspace{.1pt}\discretionary{.}{%
}{.}\hspace{.4pt}3376873}}}


\bibitem{lai2019human}
V.~Lai and C.~Tan.
\newblock On human predictions with explanations and predictions of machine learning models: A case study on deception detection.
\newblock In {\em Proceedings of the conference on fairness, accountability, and transparency}, pp. 29--38, 2019. \href{https://doi.org/10.1145/3287560.3287590}
{doi: {{%
10\hspace{.1pt}\discretionary{.}{%
}{.}\hspace{.4pt}1145\discretionary{/}{%
}{/}3287560\hspace{.1pt}\discretionary{.}{%
}{.}\hspace{.4pt}3287590}}}


\bibitem{lee2020biobert}
J.~Lee, W.~Yoon, S.~Kim, D.~Kim, S.~Kim, C.~H. So, and J.~Kang.
\newblock Biobert: a pre-trained biomedical language representation model for biomedical text mining.
\newblock {\em Bioinformatics}, 36(4):1234--1240, 2020. \href{https://doi.org/10.1093/bioinformatics/btz682}
{doi: {{%
10\hspace{.1pt}\discretionary{.}{%
}{.}\hspace{.4pt}1093\discretionary{/}{%
}{/}bioinformatics\discretionary{/}{%
}{/}btz682}}}


\bibitem{li2023latent}
J.~Li, D.~Tang, X.~Lu, F.~Sun, K.~Jiang, and J.~Ruan.
\newblock Latent space feature representation on multiple biological network for synthetic lethality interaction prediction.
\newblock In {\em 2023 IEEE International Conference on Bioinformatics and Biomedicine (BIBM)}, pp. 1236--1241, 2023. \href{https://doi.org/10.1109/BIBM58861.2023.10385727}
{doi: {{%
10\hspace{.1pt}\discretionary{.}{%
}{.}\hspace{.4pt}1109\discretionary{/}{%
}{/}BIBM58861\hspace{.1pt}\discretionary{.}{%
}{.}\hspace{.4pt}2023\hspace{.1pt}\discretionary{.}{%
}{.}\hspace{.4pt}10385727}}}


\bibitem{lim1997empirical}
K.~H. Lim, L.~M. Ward, and I.~Benbasat.
\newblock An empirical study of computer system learning: Comparison of co-discovery and self-discovery methods.
\newblock {\em Information Systems Research}, 8(3):254--272, 1997. \href{https://doi.org/10.1287/isre.8.3.254}
{doi: {{%
10\hspace{.1pt}\discretionary{.}{%
}{.}\hspace{.4pt}1287\discretionary{/}{%
}{/}isre\hspace{.1pt}\discretionary{.}{%
}{.}\hspace{.4pt}8\hspace{.1pt}\discretionary{.}{%
}{.}\hspace{.4pt}3\hspace{.1pt}\discretionary{.}{%
}{.}\hspace{.4pt}254}}}


\bibitem{linderman2019clustering}
G.~C. Linderman and S.~Steinerberger.
\newblock Clustering with t-sne, provably.
\newblock {\em SIAM journal on mathematics of data science}, 1(2):313--332, 2019. \href{https://doi.org/10.1137/18M1216134}
{doi: {{%
10\hspace{.1pt}\discretionary{.}{%
}{.}\hspace{.4pt}1137\discretionary{/}{%
}{/}18M1216134}}}


\bibitem{liu2016EgoComp}
D.~Liu, F.~Guo, B.~Deng, H.~Qu, and Y.~Wu.
\newblock Egocomp: A node-link-based technique for visual comparison of ego-networks.
\newblock {\em Information Visualization}, 16, 09 2016. \href{https://doi.org/10.1177/1473871616667632}
{doi: {{%
10\hspace{.1pt}\discretionary{.}{%
}{.}\hspace{.4pt}1177\discretionary{/}{%
}{/}1473871616667632}}}


\bibitem{liu2018synthetic}
L.~Liu, X.~Chen, C.~Hu, D.~Zhang, Z.~Shao, Q.~Jin, J.~Yang, H.~Xie, B.~Liu, M.~Hu, et~al.
\newblock Synthetic lethality-based identification of targets for anticancer drugs in the human signaling network.
\newblock {\em Scientific Reports}, 8(1):8440, 2018. \href{https://doi.org/10.1038/s41598-018-26783-w}
{doi: {{%
10\hspace{.1pt}\discretionary{.}{%
}{.}\hspace{.4pt}1038\discretionary{/}{%
}{/}s41598\discretionary{%
}{-}{-}018\discretionary{%
}{-}{-}26783\discretionary{%
}{-}{-}w}}}


\bibitem{liu2022corgie}
Z.~Liu, Y.~Wang, J.~Bernard, and T.~Munzner.
\newblock Visualizing graph neural networks with corgie: Corresponding a graph to its embedding.
\newblock {\em IEEE Transactions on Visualization and Computer Graphics}, 28(6):2500--2516, 2022. \href{https://doi.org/10.1109/TVCG.2022.3148197}
{doi: {{%
10\hspace{.1pt}\discretionary{.}{%
}{.}\hspace{.4pt}1109\discretionary{/}{%
}{/}TVCG\hspace{.1pt}\discretionary{.}{%
}{.}\hspace{.4pt}2022\hspace{.1pt}\discretionary{.}{%
}{.}\hspace{.4pt}3148197}}}


\bibitem{lundberg2020local}
S.~M. Lundberg, G.~Erion, H.~Chen, A.~DeGrave, J.~M. Prutkin, B.~Nair, R.~Katz, J.~Himmelfarb, N.~Bansal, and S.-I. Lee.
\newblock From local explanations to global understanding with explainable ai for trees.
\newblock {\em Nature machine intelligence}, 2(1):56--67, 2020. \href{https://doi.org/10.1038/s42256-019-0138-9}
{doi: {{%
10\hspace{.1pt}\discretionary{.}{%
}{.}\hspace{.4pt}1038\discretionary{/}{%
}{/}s42256\discretionary{%
}{-}{-}019\discretionary{%
}{-}{-}0138\discretionary{%
}{-}{-}9}}}


\bibitem{luo2020parameterized}
D.~Luo, W.~Cheng, D.~Xu, W.~Yu, B.~Zong, H.~Chen, and X.~Zhang.
\newblock Parameterized explainer for graph neural network.
\newblock {\em Advances in neural information processing systems}, 33:19620--19631, 2020. \href{https://doi.org/10.48550/arXiv.2011.04573}
{doi: {{%
10\hspace{.1pt}\discretionary{.}{%
}{.}\hspace{.4pt}48550\discretionary{/}{%
}{/}arXiv\hspace{.1pt}\discretionary{.}{%
}{.}\hspace{.4pt}2011\hspace{.1pt}\discretionary{.}{%
}{.}\hspace{.4pt}04573}}}


\bibitem{ma2023modeling}
S.~Ma, M.~Sun, and X.~Ma.
\newblock Modeling adaptive expression of robot learning engagement and exploring its effects on human teachers.
\newblock {\em ACM Transactions on Computer-Human Interaction}, 30(5):1--48, 2023. \href{https://doi.org/10.1145/3571813}
{doi: {{%
10\hspace{.1pt}\discretionary{.}{%
}{.}\hspace{.4pt}1145\discretionary{/}{%
}{/}3571813}}}


\bibitem{mackeprang2019discovering}
M.~Mackeprang, C.~M{\"u}ller-Birn, and M.~T. Stauss.
\newblock Discovering the sweet spot of human-computer configurations: A case study in information extraction.
\newblock {\em Proceedings of the ACM on Human-Computer Interaction}, 3(CSCW):1--30, 2019. \href{https://doi.org/10.1145/3359297}
{doi: {{%
10\hspace{.1pt}\discretionary{.}{%
}{.}\hspace{.4pt}1145\discretionary{/}{%
}{/}3359297}}}


\bibitem{McInnes2018UMAPUM}
L.~McInnes and J.~Healy.
\newblock Umap: Uniform manifold approximation and projection for dimension reduction.
\newblock {\em ArXiv}, abs/1802.03426, 2018. \href{https://doi.org/10.21105/joss.00861}
{doi: {{%
10\hspace{.1pt}\discretionary{.}{%
}{.}\hspace{.4pt}21105\discretionary{/}{%
}{/}joss\hspace{.1pt}\discretionary{.}{%
}{.}\hspace{.4pt}00861}}}


\bibitem{oakes2023workflow}
B.~J. Oakes, M.~Famelis, and H.~Sahraoui.
\newblock Building domain-specific machine learning workflows: A conceptual framework for the state-of-the-practice.
\newblock {\em ACM Trans. Softw. Eng. Methodol.}, 2023. \href{https://doi.org/10.1145/3638243}
{doi: {{%
10\hspace{.1pt}\discretionary{.}{%
}{.}\hspace{.4pt}1145\discretionary{/}{%
}{/}3638243}}}


\bibitem{SLC}
N.~J. O'Neil, M.~L. Bailey, and P.~Hieter.
\newblock Synthetic lethality and cancer.
\newblock {\em Nature Reviews Genetics}, 18(10):613--623, 2017. \href{https://doi.org/10.1038/nrg.2017.47}
{doi: {{%
10\hspace{.1pt}\discretionary{.}{%
}{.}\hspace{.4pt}1038\discretionary{/}{%
}{/}nrg\hspace{.1pt}\discretionary{.}{%
}{.}\hspace{.4pt}2017\hspace{.1pt}\discretionary{.}{%
}{.}\hspace{.4pt}47}}}


\bibitem{ryan2023complex}
C.~J. Ryan, L.~P.~S. Devakumar, S.~J. Pettitt, and C.~J. Lord.
\newblock Complex synthetic lethality in cancer.
\newblock {\em Nature Genetics}, 55(12):2039--2048, 2023. \href{https://doi.org/10.1038/s41588-023-01557-x}
{doi: {{%
10\hspace{.1pt}\discretionary{.}{%
}{.}\hspace{.4pt}1038\discretionary{/}{%
}{/}s41588\discretionary{%
}{-}{-}023\discretionary{%
}{-}{-}01557\discretionary{%
}{-}{-}x}}}


\bibitem{shan2021reinforcement}
C.~Shan, Y.~Shen, Y.~Zhang, X.~Li, and D.~Li.
\newblock Reinforcement learning enhanced explainer for graph neural networks.
\newblock {\em Advances in Neural Information Processing Systems}, 34:22523--22533, 2021.

\bibitem{shi2023RetroLens}
C.~Shi, Y.~Hu, S.~Wang, S.~Ma, C.~Zheng, X.~Ma, and Q.~Luo.
\newblock Retrolens: A human-ai collaborative system for multi-step retrosynthetic route planning.
\newblock In {\em Proceedings of the 2023 CHI Conference on Human Factors in Computing Systems}, CHI '23,  article no. 770,  20 pages. Association for Computing Machinery, New York, NY, USA, 2023. \href{https://doi.org/10.1145/3544548.3581469}
{doi: {{%
10\hspace{.1pt}\discretionary{.}{%
}{.}\hspace{.4pt}1145\discretionary{/}{%
}{/}3544548\hspace{.1pt}\discretionary{.}{%
}{.}\hspace{.4pt}3581469}}}


\bibitem{spinelli2022meta}
I.~Spinelli, S.~Scardapane, and A.~Uncini.
\newblock A meta-learning approach for training explainable graph neural networks.
\newblock {\em IEEE Transactions on Neural Networks and Learning Systems}, 2022. \href{https://doi.org/10.1109/TNNLS.2022.3171398}
{doi: {{%
10\hspace{.1pt}\discretionary{.}{%
}{.}\hspace{.4pt}1109\discretionary{/}{%
}{/}TNNLS\hspace{.1pt}\discretionary{.}{%
}{.}\hspace{.4pt}2022\hspace{.1pt}\discretionary{.}{%
}{.}\hspace{.4pt}3171398}}}


\bibitem{topatana2020advances}
W.~Topatana, S.~Juengpanich, S.~Li, J.~Cao, J.~Hu, J.~Lee, K.~Suliyanto, D.~Ma, B.~Zhang, M.~Chen, et~al.
\newblock Advances in synthetic lethality for cancer therapy: cellular mechanism and clinical translation.
\newblock {\em Journal of hematology \& oncology}, 13:1--22, 2020. \href{https://doi.org/10.1186/s13045-020-00956-5}
{doi: {{%
10\hspace{.1pt}\discretionary{.}{%
}{.}\hspace{.4pt}1186\discretionary{/}{%
}{/}s13045\discretionary{%
}{-}{-}020\discretionary{%
}{-}{-}00956\discretionary{%
}{-}{-}5}}}


\bibitem{Maaten2008VisualizingDU}
L.~van~der Maaten and G.~E. Hinton.
\newblock Visualizing data using t-sne.
\newblock {\em Journal of Machine Learning Research}, 9:2579--2605, 2008.

\bibitem{velivckovic2017graph}
P.~Veli{\v{c}}kovi{\'c}, G.~Cucurull, A.~Casanova, A.~Romero, P.~Lio, and Y.~Bengio.
\newblock Graph attention networks.
\newblock {\em arXiv preprint arXiv:1710.10903}, 2017. \href{https://doi.org/10.48550/arXiv.1710.10903}
{doi: {{%
10\hspace{.1pt}\discretionary{.}{%
}{.}\hspace{.4pt}48550\discretionary{/}{%
}{/}arXiv\hspace{.1pt}\discretionary{.}{%
}{.}\hspace{.4pt}1710\hspace{.1pt}\discretionary{.}{%
}{.}\hspace{.4pt}10903}}}


\bibitem{Wang2022ComputationalMD}
J.~Wang, Q.~Zhang, J.~cai Han, Y.~Zhao, C.~Zhao, B.~Yan, C.~Dai, L.~Wu, Y.~Wen, Y.~Zhang, D.~Leng, Z.~Wang, X.~Yang, S.~He, and X.~Bo.
\newblock Computational methods, databases and tools for synthetic lethality prediction.
\newblock {\em Briefings in Bioinformatics}, 23, 2022. \href{https://doi.org/10.1093/bib/bbac106}
{doi: {{%
10\hspace{.1pt}\discretionary{.}{%
}{.}\hspace{.4pt}1093\discretionary{/}{%
}{/}bib\discretionary{/}{%
}{/}bbac106}}}


\bibitem{Wang2023User-Centric}
Q.~Wang, K.~Huang, P.~Chandak, M.~Zitnik, and N.~Gehlenborg.
\newblock Extending the nested model for user-centric xai: A design study on gnn-based drug repurposing.
\newblock {\em IEEE Transactions on Visualization and Computer Graphics}, 29(1):1266--1276, 2023. \href{https://doi.org/10.1109/TVCG.2022.3209435}
{doi: {{%
10\hspace{.1pt}\discretionary{.}{%
}{.}\hspace{.4pt}1109\discretionary{/}{%
}{/}TVCG\hspace{.1pt}\discretionary{.}{%
}{.}\hspace{.4pt}2022\hspace{.1pt}\discretionary{.}{%
}{.}\hspace{.4pt}3209435}}}


\bibitem{wang2022NSF4SL}
S.~Wang, Y.~Feng, X.~Liu, Y.~Liu, M.~Wu, and J.~Zheng.
\newblock Nsf4sl: negative-sample-free contrastive learning for ranking synthetic lethal partner genes in human cancers.
\newblock {\em Bioinformatics (Oxford, England)}, 38:ii13--ii19, 09 2022. \href{https://doi.org/10.1093/bioinformatics/btac462}
{doi: {{%
10\hspace{.1pt}\discretionary{.}{%
}{.}\hspace{.4pt}1093\discretionary{/}{%
}{/}bioinformatics\discretionary{/}{%
}{/}btac462}}}


\bibitem{wang2021kg4sl}
S.~Wang, F.~Xu, Y.~Li, J.~Wang, K.~Zhang, Y.~Liu, M.~Wu, and J.~Zheng.
\newblock Kg4sl: knowledge graph neural network for synthetic lethality prediction in human cancers.
\newblock {\em Bioinformatics}, 37(Supplement\_1):i418--i425, 2021. \href{https://doi.org/10.1093/bioinformatics/btab271}
{doi: {{%
10\hspace{.1pt}\discretionary{.}{%
}{.}\hspace{.4pt}1093\discretionary{/}{%
}{/}bioinformatics\discretionary{/}{%
}{/}btab271}}}


\bibitem{wang2013identification}
X.~Wang and R.~Simon.
\newblock Identification of potential synthetic lethal genes to p53 using a computational biology approach.
\newblock {\em BMC medical genomics}, 6(1):1--10, 2013. \href{https://doi.org/10.1186/1755-8794-6-30}
{doi: {{%
10\hspace{.1pt}\discretionary{.}{%
}{.}\hspace{.4pt}1186\discretionary{/}{%
}{/}1755\discretionary{%
}{-}{-}8794\discretionary{%
}{-}{-}6\discretionary{%
}{-}{-}30}}}


\bibitem{wang2022Explanations}
X.~Wang and M.~Yin.
\newblock Effects of explanations in ai-assisted decision making: Principles and comparisons.
\newblock {\em ACM Trans. Interact. Intell. Syst.}, 12(4),  article no. 27,  36 pages, nov 2022. \href{https://doi.org/10.1145/3519266}
{doi: {{%
10\hspace{.1pt}\discretionary{.}{%
}{.}\hspace{.4pt}1145\discretionary{/}{%
}{/}3519266}}}


\bibitem{wold1987principal}
S.~Wold, K.~Esbensen, and P.~Geladi.
\newblock Principal component analysis.
\newblock {\em Chemometrics and intelligent laboratory systems}, 2(1-3):37--52, 1987. \href{https://doi.org/10.1016/0169-7439(87)80084-9}
{doi: {{%
10\hspace{.1pt}\discretionary{.}{%
}{.}\hspace{.4pt}1016\discretionary{/}{%
}{/}0169\discretionary{%
}{-}{-}7439\discretionary{%
}{(}{(}87\discretionary{)}{%
}{)}80084\discretionary{%
}{-}{-}9}}}


\bibitem{wu2021synthetic}
L.-L. Wu, Y.-Q. Wen, X.-X. Yang, B.-W. Yan, S.~He, and X.-C. Bo.
\newblock Synthetic lethal interactions prediction based on multiple similarity measures fusion.
\newblock {\em Journal of Computer Science and Technology}, 36:261--275, 2021. \href{https://doi.org/10.1007/s11390-021-0866-2}
{doi: {{%
10\hspace{.1pt}\discretionary{.}{%
}{.}\hspace{.4pt}1007\discretionary{/}{%
}{/}s11390\discretionary{%
}{-}{-}021\discretionary{%
}{-}{-}0866\discretionary{%
}{-}{-}2}}}


\bibitem{wu2022survey}
X.~Wu, L.~Xiao, Y.~Sun, J.~Zhang, T.~Ma, and L.~He.
\newblock A survey of human-in-the-loop for machine learning.
\newblock {\em Future Generation Computer Systems}, 135:364--381, 2022. \href{https://doi.org/10.1016/j.future.2022.05.014}
{doi: {{%
10\hspace{.1pt}\discretionary{.}{%
}{.}\hspace{.4pt}1016\discretionary{/}{%
}{/}j\hspace{.1pt}\discretionary{.}{%
}{.}\hspace{.4pt}future\hspace{.1pt}\discretionary{.}{%
}{.}\hspace{.4pt}2022\hspace{.1pt}\discretionary{.}{%
}{.}\hspace{.4pt}05\hspace{.1pt}\discretionary{.}{%
}{.}\hspace{.4pt}014}}}


\bibitem{yang2021mapping}
C.~Yang, Y.~Guo, R.~Qian, Y.~Huang, L.~Zhang, J.~Wang, X.~Huang, Z.~Liu, W.~Qin, C.~Wang, et~al.
\newblock Mapping the landscape of synthetic lethal interactions in liver cancer.
\newblock {\em Theranostics}, 11(18):9038, 2021. \href{https://doi.org/10.7150/thno.63416}
{doi: {{%
10\hspace{.1pt}\discretionary{.}{%
}{.}\hspace{.4pt}7150\discretionary{/}{%
}{/}thno\hspace{.1pt}\discretionary{.}{%
}{.}\hspace{.4pt}63416}}}


\bibitem{yang2012genomics}
W.~Yang, J.~Soares, P.~Greninger, E.~J. Edelman, H.~Lightfoot, S.~Forbes, N.~Bindal, D.~Beare, J.~A. Smith, I.~R. Thompson, et~al.
\newblock Genomics of drug sensitivity in cancer (gdsc): a resource for therapeutic biomarker discovery in cancer cells.
\newblock {\em Nucleic acids research}, 41(D1):D955--D961, 2012. \href{https://doi.org/10.1093/nar/gks1111}
{doi: {{%
10\hspace{.1pt}\discretionary{.}{%
}{.}\hspace{.4pt}1093\discretionary{/}{%
}{/}nar\discretionary{/}{%
}{/}gks1111}}}


\bibitem{ying2019gnnexplainer}
Z.~Ying, D.~Bourgeois, J.~You, M.~Zitnik, and J.~Leskovec.
\newblock Gnnexplainer: Generating explanations for graph neural networks.
\newblock {\em Advances in neural information processing systems}, 32, 2019. \href{https://doi.org/10.48550/arXiv.1903.03894}
{doi: {{%
10\hspace{.1pt}\discretionary{.}{%
}{.}\hspace{.4pt}48550\discretionary{/}{%
}{/}arXiv\hspace{.1pt}\discretionary{.}{%
}{.}\hspace{.4pt}1903\hspace{.1pt}\discretionary{.}{%
}{.}\hspace{.4pt}03894}}}


\bibitem{yu2023improved}
X.~Yu, D.~Liang, and Q.~Li.
\newblock Improved graphsvx for gnn explanations based on cross entropy.
\newblock In {\em 2023 4th International Conference on Electronic Communication and Artificial Intelligence (ICECAI)}, pp. 147--152. IEEE, 2023. \href{https://doi.org/10.1109/ICECAI58670.2023.10176786}
{doi: {{%
10\hspace{.1pt}\discretionary{.}{%
}{.}\hspace{.4pt}1109\discretionary{/}{%
}{/}ICECAI58670\hspace{.1pt}\discretionary{.}{%
}{.}\hspace{.4pt}2023\hspace{.1pt}\discretionary{.}{%
}{.}\hspace{.4pt}10176786}}}


\bibitem{yuan2021explainability}
H.~Yuan, H.~Yu, J.~Wang, K.~Li, and S.~Ji.
\newblock On explainability of graph neural networks via subgraph explorations.
\newblock In {\em International conference on machine learning}, pp. 12241--12252. PMLR, 2021. \href{https://doi.org/10.48550/arXiv.2102.05152}
{doi: {{%
10\hspace{.1pt}\discretionary{.}{%
}{.}\hspace{.4pt}48550\discretionary{/}{%
}{/}arXiv\hspace{.1pt}\discretionary{.}{%
}{.}\hspace{.4pt}2102\hspace{.1pt}\discretionary{.}{%
}{.}\hspace{.4pt}05152}}}


\bibitem{zhang2023mixupexplainer}
J.~Zhang, D.~Luo, and H.~Wei.
\newblock Mixupexplainer: Generalizing explanations for graph neural networks with data augmentation.
\newblock In {\em Proceedings of the 29th ACM SIGKDD Conference on Knowledge Discovery and Data Mining}, pp. 3286--3296, 2023. \href{https://doi.org/10.1145/3580305.3599435}
{doi: {{%
10\hspace{.1pt}\discretionary{.}{%
}{.}\hspace{.4pt}1145\discretionary{/}{%
}{/}3580305\hspace{.1pt}\discretionary{.}{%
}{.}\hspace{.4pt}3599435}}}


\bibitem{zhang2023kr4sl}
K.~Zhang, M.~Wu, Y.~Liu, Y.~Feng, and J.~Zheng.
\newblock Kr4sl: knowledge graph reasoning for explainable prediction of synthetic lethality.
\newblock {\em Bioinformatics}, 39(Supplement\_1):i158--i167, 2023. \href{https://doi.org/10.1093/bioinformatics/btad261}
{doi: {{%
10\hspace{.1pt}\discretionary{.}{%
}{.}\hspace{.4pt}1093\discretionary{/}{%
}{/}bioinformatics\discretionary{/}{%
}{/}btad261}}}


\bibitem{zhang2022ontoprotein}
N.~Zhang, Z.~Bi, X.~Liang, S.~Cheng, H.~Hong, S.~Deng, J.~Lian, Q.~Zhang, and H.~Chen.
\newblock Ontoprotein: Protein pretraining with gene ontology embedding.
\newblock {\em arXiv preprint arXiv:2201.11147}, 2022. \href{https://doi.org/10.48550/arXiv.2201.11147}
{doi: {{%
10\hspace{.1pt}\discretionary{.}{%
}{.}\hspace{.4pt}48550\discretionary{/}{%
}{/}arXiv\hspace{.1pt}\discretionary{.}{%
}{.}\hspace{.4pt}2201\hspace{.1pt}\discretionary{.}{%
}{.}\hspace{.4pt}11147}}}


\bibitem{zhang2020AI-Assisted}
Y.~Zhang, Q.~V. Liao, and R.~K.~E. Bellamy.
\newblock Effect of confidence and explanation on accuracy and trust calibration in ai-assisted decision making.
\newblock In {\em Proceedings of the 2020 Conference on Fairness, Accountability, and Transparency}, FAT* '20,  11 pages, p. 295–305. Association for Computing Machinery, New York, NY, USA, 2020. \href{https://doi.org/10.1145/3351095.3372852}
{doi: {{%
10\hspace{.1pt}\discretionary{.}{%
}{.}\hspace{.4pt}1145\discretionary{/}{%
}{/}3351095\hspace{.1pt}\discretionary{.}{%
}{.}\hspace{.4pt}3372852}}}


\bibitem{zhang2023emerging}
Y.~Zhang, Q.~Yao, L.~Yue, X.~Wu, Z.~Zhang, Z.~Lin, and Y.~Zheng.
\newblock Emerging drug interaction prediction enabled by a flow-based graph neural network with biomedical network.
\newblock {\em Nature Computational Science}, 3(12):1023--1033, 2023. \href{https://doi.org/10.1038/s43588-023-00558-4}
{doi: {{%
10\hspace{.1pt}\discretionary{.}{%
}{.}\hspace{.4pt}1038\discretionary{/}{%
}{/}s43588\discretionary{%
}{-}{-}023\discretionary{%
}{-}{-}00558\discretionary{%
}{-}{-}4}}}


\bibitem{zhao2022Human-in-the-loop}
Z.~Zhao, P.~Xu, C.~Scheidegger, and L.~Ren.
\newblock Human-in-the-loop extraction of interpretable concepts in deep learning models.
\newblock {\em IEEE Transactions on Visualization and Computer Graphics}, 28(1):780--790, 2022. \href{https://doi.org/10.1109/TVCG.2021.3114837}
{doi: {{%
10\hspace{.1pt}\discretionary{.}{%
}{.}\hspace{.4pt}1109\discretionary{/}{%
}{/}TVCG\hspace{.1pt}\discretionary{.}{%
}{.}\hspace{.4pt}2021\hspace{.1pt}\discretionary{.}{%
}{.}\hspace{.4pt}3114837}}}


\bibitem{zhu2023slgnn}
Y.~Zhu, Y.~Zhou, Y.~Liu, X.~Wang, and J.~Li.
\newblock {SLGNN: synthetic lethality prediction in human cancers based on factor-aware knowledge graph neural network}.
\newblock {\em Bioinformatics}, 39(2):btad015, 01 2023. \href{https://doi.org/10.1093/bioinformatics/btad015}
{doi: {{%
10\hspace{.1pt}\discretionary{.}{%
}{.}\hspace{.4pt}1093\discretionary{/}{%
}{/}bioinformatics\discretionary{/}{%
}{/}btad015}}}


\bibitem{zuckerman2022tangible}
O.~Zuckerman, V.~S. Press, E.~Barda, B.~Megidish, and H.~Erel.
\newblock Tangible collaboration: A human-centered approach for sharing control with an actuated-interface.
\newblock In {\em Proceedings of the 2022 CHI Conference on Human Factors in Computing Systems}, pp. 1--13, 2022. \href{https://doi.org/10.1145/3491102.3517449}
{doi: {{%
10\hspace{.1pt}\discretionary{.}{%
}{.}\hspace{.4pt}1145\discretionary{/}{%
}{/}3491102\hspace{.1pt}\discretionary{.}{%
}{.}\hspace{.4pt}3517449}}}


\end{thebibliography}

\appendix 

\end{document}